# Polarization states of gravitational waves detected by LIGO-Virgo antennas

Graduation Project
Master Thesis

Liudmila Fesik

Saint Petersburg State University

Scientific supervisor
Dr.Sci. (Phys.-Math.), leading researcher Yurij Baryshev

Reviewer
Dr.Sci. (Phys.-Math.), professor Valentin Rudenko

Saint Petersburg
2017

# Abstract

The detection of the first gravitational wave events by the Advanced LIGO Scientific Collaboration has opened a new possibility for the study of fundamental physics of gravitational interaction.

This work conducts an analysis of possible polarization states of gravitational waves (GW) radiated by the most promising types of sources to be detected by the modern interferometric antennas: coalescing compact binaries and collapsing supernovae. Several theoretical approaches to the current as well as future gravitational wave signals interpretation are discussed together with a strategy for a search for corresponding transients by means of multimessenger astronomy.

One of the aims of this thesis is to develop a new method for GW source localization depending on a polarization state of an incoming GW in the case of a detection by two interferometric antennas. Additionally, there is elaborated a further method focused on a possibility to recognise different polarization states of a GW detected by means of a network with three and more antennas in operation. The both proposed methods have been applied to the LIGO events GW150914, GW151226 and LVT151012, together with the matching of the results with the currently known electromagnetic follow-ups to these GW events.

The conclusion of this research is that there are opportunities for verifying different predictions of the scalar-tensor theories of gravitation by means of the analysis of the polarization states of the detected gravitational waves.

All in all, this work provides a new test of the theoretical assumptions about the nature of gravity as well as about the processes in relativistic compact objects of various types.

# Contents

# Chapter 1

# Introduction

Detection of gravitational wave signals by means of the Advanced LIGO antennas at the end of 2015 opened a new era in the study of the Universe by methods of gravitational-wave astronomy (Abbott et al. (2016c). Experimental confirmation of the existence of gravitational waves (hereafter GW) is of great importance for relativistic astrophysics related to understanding the physics of relativistic compact objects, massive supernovae explosions, gamma bursts and activity of galactic nuclei. With this, the new possibilities appear for testing the theory of gravity, which is also important for constructing a theory unifying fundamental physical interactions.

The modern theories of gravitation predict the existence of different polarization states of GW. Therefore, the observations of GWs can be applied for testing the theory of gravitation as well as for analysing the new interpretations of physical processes occurring under conditions of strong gravitational fields.

## 1.1 Gravitational waves in the modern theories of gravitation

The existence of gravitational waves as a free gravitational field propagating in the spacetime with the speed of light was originally predicted in the Poincarè's work (1905) (Poincarè (1905)), in which he proceeded from the analogy of the Coulomb's law with the Newton's law (as solutions of the Laplace's equation) and their relativistic generalization to the wave equation.

The theoretical description of GW radiation was firstly given by A. Einstein in the so-called "geometrical" approach to the study of gravitation in the frame of the theory of general relativity (hereafter GRT or GR) (Einstein (1916), Einstein (1918)). The basic principles of the GRT are the principle of equivalence and the principle of geometrization, which exclude gravity as an ordinary physical force field in space and reduce it to the curvature of the spacetime itself. Thereby, the property of "gravity" is attributed to the spacetime itself but not to the physical field in it. However, the refusal of homogeneity and isotropy of the spacetime leads to difficulties with the determination of the energy of the gravitational field.

The quantity characterising the energy-momentum of the gravitational field in GRT, the so-called Landau–Lifshitz pseudotensor (Landau and Lifshitz (1988)), is not a tensor of the Riemannian space since it does not preserved under common coordinate transformations. In the framework of the GRT, this means non-localizability of the energy-momentum of the gravitational field itself (Landau and Lifshitz (1988), §96, p. 362; Misner et al. (1973) §20.4, p. 467 and §35.7, p. 955). Such a property of the stress-energy-momentum pseudotensor becomes critically important in the question about the transfer of energy by a wave. In other words, if a GW is detected, its energy-momentum should be localized, i.e. transmitted to the detector. To solve this problem in the GRT, there is used a "linearized" theory in the weak-field approximation applicable in the flat Minkowski spacetime, for more details see Ch. (2.1).

Another solution to the problem of the energy of a gravitational field is proposed in the "field" approach, with a strategy developed by R. Feynman in the 1960s (Feynman (1971), Feynman et al. (1995)). In the field gravitation theory (hereafter FGT, or so-called "gravidynamics"), there is considered a localizable tensor gravitational field acting in the Minkowski spacetime.

The Feynman's field approach assumes the construction of a quantum relativistic gravitation theory based on the Lagrangian formalism in the flat spacetime, similarly to the other theories of the fundamental interactions (electromagnetic, weak and strong). Instead of the equivalence principle in the GRT, the principle of universality of gravitational interaction is the initial one in the FGT framework. It means that the interaction of a gravitational field (hereafter GF) with any other physical field occurs via its energy-momentum tensor (hereafter EMT). Other cornerstone of the FGT is the assumption of homogeneity and isotropy of the Minkowski spacetime, which, according to the Noether's theorem, guarantees the existence of the energy-momentum conservation laws. Therefore, there are no difficulties in the determination of the GF energy and the conservation laws.

The study of GWs is mainly conducted within the framework of a geometric approach based on the GRT as the most elaborated relativistic theory of gravitation. Nevertheless, at the present time, the great attention is also paid to various GR modifications, including the field approach (see reviews Clifton et al. (2012), Will (2014), Baryshev (2017)). Therefore, the actual task now is the development of new tests of the theory of gravitation, in particular, based on the analysis of observations of gravitational waves.

According to the GRT, there are only two polarization states to exist: tensor transverse "plus" and "cross" modes. Currently, there are, however, other theories being modifications of the GRT – scalar-tensor metric theories such as Brans-Dicke theory (hereafter BDT), which predict apart from tensor wave the existence of scalar transverse one. On the other hand, within the framework of the FGT, there are possible scalar longitudinal waves generated by change of the EMT trace of a source.

It is important to note the principal difference between the FGT and scalar-tensor modifications of the GRT. In particular, in the frame of the BDT, there is introduced an external scalar field $\phi_{\mathrm{BD}}$ having a coupling constant $\omega$, while in the FGT the scalar field is the natural internal part of the symmetric tensor field $\psi^{ik}$, i.e. its trace $\psi$, with the Newtonian gravitational coupling constant $G$.

The existence of other polarization states, namely the scalar transverse and longitudinal waves different from the tensor modes considered in GR, can be verified by analysis of GW signals recorded by the aLIGO and aVirgo antennas.

In this work, a method has been developed that makes it possible to recognise tensor and scalar GW polarization states by means of observations on the existing interferometric antennas and, consequently, to test the theoretical assumptions about the nature of gravity as well as about the processes in the relativistic compact objects of various types. There is shown the application of the method to the GW events discovered in 2015 together with the analysis of follow-ups events in the electromagnetic branch, as well as the possibility of applying this method in the future experiments/observations.

## 1.2 Possible sources of GW radiation

Despite the fact that GW radiation was theoretically predicted by A. Einstein as far back as 1916, the question about a practical detection of GWs remained open for decades (Rudenko, V. N. (2017)).

The first indirect confirmation of the existence of GW radiation was the discovery of a binary system with the pulsar PSR1913+16 in 1974, and its further observations by J. Taylor and R. Hulse (Weisberg and Taylor (2005)). The system PSR1913 + 16 is a neutron star binary with one of the stars in the pair – a pulsar, its radio pulses are used as a clock to monitor the orbital period of the system. According to the relativistic theory of gravity (either the GR or the FGT), two stars rotating around each other lose the energy as a result of GW radiation, and consequently, the radius of their orbit and the orbital period should decrease.

During more than 40 years after this discovery, the observations of the PSR1913+16 and their analysis had been conducted (Weisberg and Huang (2016)), which have showed that the systematic decrease in the orbital period of the system is very accurately consistent with the theoretical predictions on the energy loss due to the tensor gravitational radiation. Thereby, the existence of the GW radiation, which carries energy from the binary system, was indirectly proved. This work of J. Taylor and R. Hulse was awarded the Nobel Prize in 1993. This discovery pushed the researchers to accelerate the implementation of the most sensitive GW detectors in order to obtain the direct evidence of the existence of GWs (see reviews Rudenko, V. N. (2017).

The GW radiation should be powerful enough to give the amplitude of a GW necessary for the detection by an antenna. Given the current sensitivity of the modern interferometric antennas, the most promising for being detectable are GWs from compact binary coalescence (CBC) and core-collapse supernovae (CCSN). The possibility of detecting waves from each type of source depends on the energy radiated in GWs, the distance to the object, the pulse duration and the frequency. As well as the frequency of such events at the considered distance is also important.

Let us consider main sources of the expected GW events. Note that under the relativistic compact objects (hereafter RCO) we consider a class of various astrophysical objects which possible within the framework of modern gravity theories and which have dimensions close to the gravitational radius $R_G = GM/c^2$. Thereby, the term "RCO" includes both the black holes in GR (hereafter BH) and, also, possible in FGT physical objects with definite surfaces, measurable magnetic fields as well as with finite gravitational forces without horizons and singularities for any masses (Sokolov and Zharykov (1993), Sokolov (2015), Baryshev (2017)).

Compact binary coalescence (CBC) is a class of GW sources with two RCOs on a common orbit. According to the adopted terminology, such pairs can be either neutron stars (hereafter NS-NS), or a neutron star with a massive RCO (NS-RCO, or NS-BH with a "black hole" in the GR frame), as well as a couple of two RCO (RCO-RCO, BH-BH). Gravitational radiation during the orbital motion of such objects "takes away" from the system both the energy and the angular momentum, which causes a decrease in orbital radius up to the merger into one RCO.

This class of GW sources is of particular interest because the stage of the orbital motion just prior to the merger – the so-called "inspiral" phase can be accurately modelled, which makes the predictions about the waveform and the frequency of a GW signal depending on the masses of the incoming objects. Therefore, observations of CBC might give an excellent test for the verification of gravitation theories. And in the case of identification of a GW event with its counterpart in the EM branch, this also makes it possible to determine the position of the GW source and the distance to it precisely. In the corresponding section (Ch. (2.2)) it will be considered what kind of information about the parameters of a CBC can be obtained from the observed GW signal.

Another type of GW events is connected with the explosions of massive supernovae. GW radiation arises as a result of the gravitational collapse of the degenerate core of the star in the late stages of its evolution, resulting in the formation of a compact object such as a neutron star or RCO. In this case, a huge amount of energy is released, of the order of $M_\odot c^2$, most of which is carried away by neutrinos and some (still undetermined) portion – by GWs.

An important scenario of the core-collapse of a massive supernova was proposed in the works of Imshennik and Nadezhin (Imshennik (2010)): due to a strong rotation of the core, there firstly occurs the formation of an RCO binary radiating tensor GWs, and then merging into a single RCO with the possible scalar GW radiation.

Although supernovae may be a powerful source of gravitational radiation, up to now there are many uncertainties in the modelling of the collapse mechanism itself. So it is difficult to make sufficiently reliable assumptions about the amplitude and the waveform of a GW from a supernovae (Thorne (1989), Maggiore (2006), Coccia et al. (2004), Burrows (2013)). Massive supernovae can differ greatly in the nature of the processes occurring in them but for the purposes of GW study, the SN bursts are divided into two types: those resulting from an asymmetric collapse of the core (in the GR) and others as a result of a spherically symmetric core-collapse (in the scalar-tensor theories of gravitation). In addition, the speed of rotation and the presence of a magnetic field should be taken into account.

The importance of a separate consideration of these types lies in the fact that according to the GR, tensor waves can arise only from an asymmetric collapse (Misner et al. (1973)), whereas both the scalar-tensor metric theories and the FGT predict the existence of a scalar GW mode, which may occur as a result of a spherically-symmetric core-collapse (CCSN) (Novak and Ibanez (2000), Maggiore and Nicolis (2000), Coccia et al. (2004), Maggiore (2006), Baryshev, Yu. V. (1990), Baryshev (2017)).

The modern theories of CCSN make it possible to explain the stages of the evolution of a massive star before and after an explosion but there is still no theory that would explain accuratelly the relativistic collapse stage itself in order to calculate the energy of GW radiation and the observed waveform (see eg. discussion Imshennik (2010), Burrows (2013)). This uncertainty motivates the further studying of the detected GW events from the point of view of the possible origin of such a signal from a collapsing supernova such as CCSN, which will be considered in Ch. 2.3.

## 1.3 Search for the follow-ups

The detection of transients (also called "follow-ups", "counterparts", "transients") in the electromagnetic (hereafter EM) branch accompanying GW signal is of fundamental importance in the analysis of the GWs physics. Firstly, the identification of the detected GW signal with an EM counterpart will increase the confidence that there has occurred a real astrophysical event. Secondly, the joint GW and EM observations complement each other significantly in the understanding of the causing physical processes. The form of a GW signal as well as its frequency, amplitude and polarization state may provide the specific information about the mass motions necessary for the source simulation. While the identification of the GW signal with an EM transient gives it possible to estimate the physical parameters of the environment surrounding the RCO, as well as to localize the source on the sky with the calculation of the distance to it.

According to the GRT, the black holes coalescence in the vacuum does not produce any EM radiation. The same is true for the case of such a merger in the interstellar medium, where the gas density and the magnitude of the magnetic fields are too small to give a noticeable EM "follow-up". However, the RCOs coalescence in clusters, dense molecular clouds and in galactic centres may have some peculiarities in the EM spectrum due to the interaction with gas and magnetic fields.

In the case when a CBC includes at least one neutron star (or an RCO without the events horizon Sokolov and Zharykov (1993), Sokolov (2015), Baryshev (2017)), it can produce the EM radiation in a wide range of wavelengths and on different time scales. Thus, a number of studies has shown (Piran (2004), Nakar (2007)) that there may be expected the short-hard gamma-ray bursts (hereafter SGRBs) with the duration of 2 seconds or less from NS-NS and NS-BH CBCs. In the review (Lipunov and Panchenko (1996)) has been discussed that there may present short radio or optical non-thermal radiation from CBCs including at least one magnetic NS.

Another class of objects expected to give the GW radiation is the CCSN, which may produce the long-soft gamma-ray bursts (LGRBs) (Woosley (1993), MacFadyen and Woosley (1999), Piran (2004)).

The analysis of the follow-ups searches will be given in Ch. (4).

## 1.4 Task statement

1. To conduct an analysis of polarization states in the case of GW radiation by:
   (a) a compact binary coalescence (CBC);
   (b) a core-collapse supernova (CCSN).

2. To consider separately the case of scalar GW radiation from a CCSN in the framework of the field gravitation theory (FGT). To obtain a relationship between physical parameters of a pulsating CCSN and observed values of a GW signal.

3. To develop a method analysing the observational data obtained by means of interferometric antennas in order to recognize polarization states of the incoming GWs. To investigate the localization capability of GW sources in the case of detection by means of two and three interferometric antennas.

4. To apply the obtained theoretical results to GW events recorded by LIGO antennas in 2015: GW150914, GW151226 and LVT151012.

5. To analyse the information about follow-ups events in the electromagnetic spectral branch to the GW150914 and GW151226, and give recommendations for the further search for possible EM transients.

# Chapter 2

# Gravitational radiation from relativistic compact objects

In this chapter, I will consider GW radiation from the most promising objects to be detected by the modern interferometers such as LIGO, Virgo: compact binaries and core-collapse supernova (hereafter CCSN). There will be compared amplitudes of the strain given by tensor and scalar gravitational radiation from these kinds of objects in the frame of GRT as well as Feynman's field approach. At the end of the chapter, there will be given analysis of the LIGO GW events.

## 2.1 Comparison between the theoretical predictions by the GR and the FGT

In practice, the study of gravitational wave is carried out far away from the GW source where the gravitational field is weak enough so that the space is nearly flat. In this case, there can be used linearized theory in the frame of GRT, and the equations of the gravitational field can be solved as in the flat spacetime. Such an approximation is consistent with the general approach of the field theory of gravity which is based on the Minkowski spacetime (see Feynman (1971), Sokolov and Baryshev (1980), review Baryshev (2017)). Therefore, the conclusions about the GW radiation in these two approaches are mostly the same, aside from the existence of scalar GW radiation in the frame of FGT.

### 2.1.1 GW radiation in the GR. Tensor waves

For the quantitative description of a GW in the linearized GRT, an entity $h_{\mu\nu}$ is introduced, which describes a small perturbation of the flat spacetime (Minkowski) metric $\eta_{\mu\nu}$:

$$g_{\mu\nu} = \eta_{\mu\nu} + h_{\mu\nu}, \ |h_{\mu\nu}| \ll 1 \tag{2.1}$$

where $g_{\mu\nu}$ is a metric tensor of the Riemannian curved spacetime. There should be noted that in the linearized theory $h_{\mu\nu}$ is not a tensor of the Minkowski spacetime (see e.g. Schutz and Ricci (2010), Baryshev (2017)), but it is proposed to act as a tensor in the flat spacetime with the metric $\eta_{\mu\nu}$.

In the linearized GRT, the metric perturbation $h_{\mu\nu}$ is reduced to the "transverse-traceless" (heareafter TT) form $h_{\mu\nu}^{TT} = h_{\mu\nu} - \frac{1}{2}\eta_{\mu\nu}h$. And after the calibration $h_{\mu\nu;\mu}^{TT} = 0$, the Einstein Field Equations in the vacuum are the wave equation:

$$\Box h_{\mu\nu}^{TT} = 0 \tag{2.2}$$

which means that a GW propagates with the speed of light (in the vacuum).

For a single plane wave propagating in the x-direction, imposing on which the both Hilbert-Lorentz and TT calibration, there remain only two nonzero values: $h_{yz}$, $h_{yy} = -h_{zz}$. Thereby, in the frame of linearized

GTR, GW are essentially transverse waves with two independent polarization states: "plus" $h_+ = h_{yz}$ and "cross" $h_\times = \frac{1}{2}(h_{yy} - h_{zz})$ (Landau and Lifshitz (1988), Misner et al. (1973)).

Then, in general terms, the wave equation:

$$h_{\mu\nu} = (h_+ \vec{e}_+ + h_\times \vec{e}_\times) \exp(-i\omega(t - x)) \tag{2.3}$$

where $\omega$ is the circular velocity of the wave with the wave vector $k^\mu$, $\vec{e}_+, \vec{e}_\times$ – the polarization tensors characterising the orts directions for "plus" and "cross" polarizations.

It is worth to mention, the main condition on the use of the considered weak-field approximation is that the wave length should be short compared with the curvature radius of the background metric, and its amplitude – small enough (Landau and Lifshitz (1988), Misner et al. (1973)).

### 2.1.2 GW radiation in the FGT. Tensor and scalar waves

According to the Feynman's field gravitation theory (FGT), gravitation field is described by a symmetric second-rank tensor $\psi^{ik}$ in the flat Minkowski spacetime (Feynman (1971), Feynman et al. (1995), Sokolov and Baryshev (1980), Baryshev (2017)). In the work (Barnes (1965)), there was shown that such a tensor can be decomposed under the Lorentz group transformations into irreducible representations of four fields: tensor, vector and two scalar. After imposing Hilbert-Lorentz calibration on the field potentials, the tensor describes a mixture of two fields: tensor (spin-2) as a traceless second rank tensor, and scalar (spin-0) as a trace of the initial tensor $\psi = \eta_{ik}\psi^{ik}$. In this sense, the FGT is a scalar-tensor gravitation theory.

The FGT is constructed by means of Lagrangian formalism (Baryshev (2017)). In this way, the gravitational field equations as well as the equations of motion are obtained. The cornerstone of the theory is the principle of universality of gravitational interaction, according to which interaction of gravitational field with any physical field is held by its energy-momentum tensor (EMT). After taking into account the conservation laws of the energy-momentum tensor of a source, the field theory predicts the existence of two dynamical fields: tensor and scalar. Therefore, in the FGT, the field potentials $\psi^{ik}$ and the EMT of sources $T^{ik}$ can be represented by the sum of tensor (spin-2, denoted "(2)") and scalar (spin-0, denoted "(0)") parts:

$$\psi^{ik} = \psi^{ik}_{(2)} + \psi^{ik}_{(0)} = (\psi^{ik} - \frac{1}{4}\eta^{ik}\psi) + \frac{1}{4}\eta^{ik}\psi \tag{2.4}$$

$$T^{ik} = T^{ik}_{(2)} + T^{ik}_{(0)} = (T^{ik} - \frac{1}{4}\eta^{ik}T) + \frac{1}{4}\eta^{ik}T \tag{2.5}$$

The field equations with the field sources are:

$$\Box\psi^{ik}_{(2)} = \frac{8\pi G}{c^2} T^{ik}_{(2)} \tag{2.6}$$

$$\Box\psi^{ik}_{(0)} = -\frac{8\pi G}{c^2} T^{ik}_{(0)} \tag{2.7}$$

The formulae show that, in the frame of FGT, there are two kinds of gravitons: spin-2 and spin-0, which both are massless particles moving at the speed of light. Besides tensor transverse gravitation field, there is also scalar longitudinal field $\psi$ generated by the trace of the EMT of a source $T = \eta_{ik}T^{ik}$ (see 2.7). Being free, these two fields are independent of each other and act simultaneously (Sokolov and Baryshev (1980), Baryshev (2017)). Consequently, as far as free gravitational fields are being considered, a scalar field is not being disappeared nor removed (in contrast to the GRT), but rather there occurs the separation of freely acting scalar and tensor fields (Sokolov and Baryshev (1980)).

The gravitational wave equations in the vacuum can be written as:

$$\Box\psi^{ik}_{(2)} = 0, \ \psi^{ik}_{,i} = 0 \tag{2.8}$$

$$\Box \psi = 0 \tag{2.9}$$

The solution of the tensor wave equation (2.8) gives the same result as the solution of $h^{\text{TT}}$ in the linearized GRT, thereby, the existence of a tensor transverse wave with "plus" or "cross" polarization modes. The principal difference between the considered approaches, geometrical and field, lies in the prediction of the existence of a scalar wave in the frame of the FGT.

### 2.1.3 Energy and amplitude of a tensor wave

For description of energy-momentum of a gravitational field in the GRT, an entity $t^{jk}$ is introduced, which does not a real tensor and therefore, was called the "pseudotensor" (Landau and Lifshitz (1988)). It means that this "pseudotensor" of energy-momentum does not represent conserved under coordinate transformations quantities (Landau and Lifshitz (1988), §96, p.362; Misner et al. (1973) §20.4, p.467 and §35.7, p.955). For instance, as has been shown by Bauer (1918), it is possible to get the nonzero $t^{jk}$ in the flat spacetime, i.e. where, according to GRT, there is no gravitational field, just as a result of a coordinate transformation. Therefore, it makes no sense to discuss a certain value of energy-momentum of the gravitational field as well as whether or not there is any gravitational energy in a definite point of the spacetime. As a consequence, according to the GRT, the energy-momentum carried by a GW cannot be localised in a region smaller than a wavelength. It is only possible to say that asome amount of energy and momentum is in a given macroscopic region of several wavelengths in size (Misner et al. (1973)).

To avoid this problem, an effective stress-energy (or Isaacson) tensor of GWs $T_{\alpha\beta}^{\text{GW}}$ was introduced in the short-wave approximation (Isaacson (1967)):

$$T_{\alpha\beta}^{\text{GW}} = \frac{c^2}{32\pi G} \left\langle h_{\mu\nu;\alpha}^{\text{TT}} h^{\text{TT}\,\mu\nu}_{\;;\beta} \right\rangle \tag{2.10}$$

where $\langle ... \rangle$ denotes the average over a spatial volume covering several wavelengths, and $h_{\mu\nu}^{\text{TT}}$ – the calibration invariant traceless-transverse part of the $h_{\mu\nu}$.

In this form, the effective stress-energy tensor $T_{\alpha\beta}^{\text{GW}}$ can be shown to be invariant under calibration transformations and conserved when observed far away from a source (Misner et al. (1973)). Thereby, this tensor can be applied to derive an energy-momentum flux transferred by a GW in the short-wave approximation.

The energy density $w$ of a plane tensor GW is obtained from the effective stress-energy tensor (2.11):

$$w_{\text{tens}} = T_{tt}^{\text{GW}} = \frac{c^2}{16\pi G} \left( \dot{h}_+^2 + \dot{h}_\times^2 \right) \tag{2.11}$$

The energy flux $F_{\text{tens}}^{\text{GW}}$ [erg/(cm$^2\cdot$ s)]:

$$F_{\text{tens}}^{\text{GW}} = \frac{dE}{dA df} = c w_{\text{tens}} = \frac{c^3}{16\pi G} \left( \dot{h}_+^2 + \dot{h}_\times^2 \right) \tag{2.12}$$

For a standard GW sinusoidal pulse (Amaldi and Pizzella (1979)), $h(t) = h_0 \cos(\omega_0 t - kx)$ propagating in the plus $x$-direction at a frequency $f_0 = 1/P_0$, with an angular velocity $\omega_0 = 2\pi f_0$, after averaging over a period of an oscillation $P_0$, the energy flux is given by:

$$\langle F_{\text{tens}}^{\text{GW}} \rangle = \frac{c^3}{16\pi G} \omega_0^2 h_0^2 \tag{2.13}$$

where $h_0$ is a dimensionless amplitude of the wave. With the pulse duration $\tau$, the average energy per unit area:

$$\langle \varepsilon_{\text{tens}}^{\text{GW}} \rangle = \langle F_{\text{tens}}^{\text{GW}} \rangle \cdot \tau = \frac{c^3}{16\pi G} \omega_0^2 h_0^2 \tau \tag{2.14}$$

Then there can be estimated the amplitude (or strain) $h_0^{\text{tens}}$ of a tensor GW from a source at a distance $r$ radiating during the time $\tau$ the energy $\Delta E \approx 4\pi r^2 \langle \varepsilon \rangle$ (Schutz and Ricci (2010)):

$$h_0^{\text{tens}} \approx \frac{1}{\pi r f_0} \left( \frac{G}{c^3} \frac{\Delta E}{\tau} \right)^{1/2} \tag{2.15}$$

### 2.1.4 Energy and amplitude of a scalar wave in the FGT

In the frame of the FGT, the energy density of a GW is obtained from the true energy-momentum tensor and, thereby, has a certain value in any point of the spacetime (Baryshev (2017)). Therefore, unlike GR, there is no problem with the GW energy localization in the FGT.

The energy density of a scalar GW in the field theory, according to (Baryshev (1999)), is:

$$w_{\text{sc}} = \frac{c^2}{32\pi G} \dot{h}_{\text{sc}}^2 \tag{2.16}$$

The average over the period energy flux carried by the scalar wave [erg/(cm$^2\cdot$ s)]:

$$\langle F_{\text{sc}} \rangle = \frac{c^3}{64\pi G} \cdot \omega_0^2 \cdot h_0^2 \tag{2.17}$$

where $h_0$ is an dimensionless amplitude of the wave. With the pulse duration $\tau$, the average energy per unit area:

$$\langle \varepsilon_{\text{sc}} \rangle = \langle F_{\text{sc}} \rangle \cdot \tau = \frac{c^3}{64\pi G} \cdot \omega_0^2 \cdot h_0^2 \cdot \tau^2 \tag{2.18}$$

The estimation of the strain by a scalar GW from a source at the distance $r$ radiating during the time $\tau$ the energy $\Delta E \approx 4\pi r^2 \langle \varepsilon \rangle$ (Schutz and Ricci (2010)):

$$h_0^{\text{sc}} \approx \frac{2}{\pi r f_0} \left( \frac{G}{c^3} \frac{\Delta E}{\tau} \right)^{1/2} \tag{2.19}$$

The obtained above estimations show that the expected amplitude (or strain) of the tensor GW (2.15) is twice as small as the average strain of a scalar GW (2.19) carried the same amount of energy. Consequently, if there exists the scalar wave radiation, it can be possible to detect GW with smaller energies from the distant objects by means of modern detectors.

## 2.2 GW from compact binary coalescence

Compact binary coalescence (hereafter CBC) is the most available astrophysical process for GW radiation analysis because it allows a fairly accurate simulation of the gravitational waveform during the so-called "inspiral" phase (Schutz and Ricci (2010)). The next stage when coalescing bodies having reached the last stable orbit quickly converge, the merger, being inaccessible to analytic description and can be modelled only by numerical methods, which include many ad hoc assumptions. Large uncertainty also exists for the rate of the CBC events.

In the two approaches considered in this paper to the study of the theory of gravitation, there is possible to exist as tensor (in both GRT and FGT) as well as scalar (in the frame of the FGT) GW radiation from coalescing binaries. However, it will be shown that the proportion of the scalar radiation in the frame of the FGT for such objects is far too small. Therefore, for the qualitative analysis it can be neglected, i.e. to consider only tensor radiation from CBC.

### 2.2.1 GW radiation from CBC

In this part will be considered the rate of the energy radiated into the tensor GW as well as into the scalar wave under the field theory approachBaryshev (1999)).

According to GRT, the luminosity of the quadrupole GW radiation (Landau and Lifshitz (1988)) can be written as:

$$L_{\text{tens}}^{\text{GW}} = \frac{G}{45\,c^5} \dddot{\hat{D}}_{jk}^2 \tag{2.20}$$

where $\hat{D}_{jk}$ is the tensor of the reduced quadrupole moment of masses (Landau and Lifshitz (1988)):

$$\hat{D}_{jk} = \int \rho(x_j x_k - \frac{1}{3}\delta_{jk}r^2)d^3x \tag{2.21}$$

For a binary system, the energy loss due to tensor gravitational radiation is obtained by averaging over an orbit period:

$$\langle \dot{E} \rangle_{\text{tens}} = -\frac{32}{5}\frac{G^4}{c^5}\frac{m_1^2 m_2^2(m_1+m_2)}{a^5(1-e^2)^{7/2}} \times$$
$$\times \left(1 + \frac{73}{24}e^2 + \frac{37}{96}e^4\right) = -L_{\text{tens}}^{\text{GW}} \tag{2.22}$$

Then the rate of loss of the angular momentum:

$$\langle \dot{L} \rangle_{\text{tens}} = -\frac{32}{5}\frac{G^4}{c^5}\frac{m_1^2 m_2^2(m_1+m_2)^{1/2}}{a^{7/2}(1-e^2)^2}\left(1 + \frac{7}{8}e^2\right) \tag{2.23}$$

There should be mention that in the case of the circular orbit, $e \equiv 0, a \equiv R$ – the radius of the orbit, the luminosity of GW quadrupole radiation:

$$L_{\text{tens}}^{\text{GW}} = -\langle \dot{E} \rangle = \frac{32}{5}\frac{G^4}{c^5}\frac{M^3\mu^2}{R^5} \tag{2.24}$$

Concerning to the scalar GW (Baryshev (1999)), there has been shown that the energy loss in a CBC is:

$$\langle \dot{E} \rangle_{\text{sc}} = -\frac{1}{4}\frac{G^4}{c^5}\frac{m_1^2 m_2^2(m_1+m_2)}{a^5(1-e^2)^{7/2}}\left(e^2 + \frac{1}{4}e^4\right) = -L_{\text{sc}}^{\text{GW}} \tag{2.25}$$

It is important to note that, in contrast to the tensor mode, the energy loss in the case of scalar GW radiation in the frame of the FGT directly depends on the value of the orbit eccentricity $e$. Thereby, the ratio of the scalar luminosity to the tensor one for the binary system is:

$$\frac{\langle \dot{E} \rangle_{\text{sc}}}{\langle \dot{E} \rangle_{\text{tens}}} = \frac{\left(e^2 + \frac{1}{4}e^4\right)}{\left(1 + \frac{73}{24}e^2 + \frac{37}{96}e^4\right)} \tag{2.26}$$

This means that there is no scalar radiation radiation for the case of the circular orbit $e \equiv 0$, and for small eccentricities it does not exceed 1% as has been shown by (Baryshev (1999)) for the case of the binary pulsar PSR 1913 + 16.

In the next part, there will be shown that the loss of energy due to GW radiation leads to an eccentricity decrease. Therefore, for binary systems in last stages of their coalescence, the orbit eccentricity can be neglected. Consequently, scalar radiation can be considered as not essential for the qualitative analysis and further, there will be focused only on the analysis of tensor radiation from CBC.

### 2.2.2 The change of radius and eccentricity due to GW radiation

The changing of the semi-major axis of the binary orbit $\dot{a}$ and the eccentricity $\dot{e}$ with the time caused by the gravitational radiation can be obtained by $\langle \dot{E} \rangle$ and $\langle \dot{L} \rangle$:

$$\langle \dot{a} \rangle = -\frac{64}{5}\frac{m_1 m_2(m_1+m_2)}{a^3(1-e^2)^{7/2}}\left(1 + \frac{73}{24}e^2 + \frac{37}{96}e^4\right) \tag{2.27}$$

$$\langle \dot{e} \rangle = -\frac{304}{15} \frac{m_1 m_2 (m_1 + m_2) e}{a^4 (1 - e^2)^{5/2}} \left(1 + \frac{121}{304} e^2\right) \tag{2.28}$$

Since the derivative $\dot{e} < 0$ the radiation reaction leads to a decrease in the eccentricity and the approach of the orbit to the circular one.

For the case of the circular orbit, $e = 0, a = R$, the change of the orbit radius $R$ with the time:

$$\frac{dR}{dt} = -\frac{64}{5} \frac{M^2 \mu}{R^3} \tag{2.29}$$

which leads to the solution for the radius:

$$R^4 = \frac{256}{5} M^2 \mu (t_c - t) \tag{2.30}$$

where the time to the coalescence $t_c$ (or merge), at which $R = 0$, can be found by assuming that, at the time of the observations start $t = 0$, the orbit radius $R = R_s$. Then:

$$t_c = \frac{5}{256} \frac{R_s^4}{M^2 \mu} \tag{2.31}$$

### 2.2.3 "Chirp"-mass and frequency

For the GW radiation description, it is convenient to introduce the "chirp"-mass $\mathfrak{M}$:

$$\mathfrak{M} = \frac{(m_1 m_2)^{3/5}}{(m_1 + m_2)^{1/5}} \tag{2.32}$$

Then, using Newtonian laws of the motion and (2.24), there can be derived the relation between the "chirp"-mass $\mathfrak{M}$, the GW frequency $f$ and its time derivative $\dot{f}$ (see eg., Abbott et al. (2016d)). After the integration, the explicit relation between the "chirp"-mass, the wave frequency and the Time to the coalescence $t_c$ can be written as:

$$f_{GW}(t) = \frac{5^{3/8}}{8\pi} \left(\frac{c^3}{G\mathfrak{M}}\right)^{5/8} (t_c - t)^{-3/8} \tag{2.33}$$

This formulae shows that with the frequency and the coalescence time known by the observations, the "chirp"-mass of the binary can be estimated directly.

### 2.2.4 The tensor waveform for CBC

The tensor waveform for the coalescing binary system on the circular orbit can be simply derived at the "inspiral" stage in the zero Post-Newtonian approximation (hereafter 0-PN). The higher orders of approximation can be found in a wide range of articles (see eg. Blanchet (2014)). In this work, I give a qualitative analysis of the GW radiation, thus there will be used only 0-PN approximation in the assumption of the almost circular orbit $e \approx 0$.

The general form of a tensor GW, "plus" (+) and "cross" (×), in 0-PN approximation can be written as:

$$h_{+,\times}(t) = \frac{2\mu x}{r} \frac{G}{c^2} H_{+,\times}(t) \tag{2.34}$$

Taking the total mass of the binary $M = m_1 + m_2$, $\mu = m_1 m_2 / M$ – its reduced mass. The time dependence is shown through the orbital phase $\psi(t)$:

$$H_+(t) = -(1 + \cos^2 i) \cos 2\psi(t) \tag{2.35a}$$

$$H_\times(t) = -2 \cos i \sin 2\psi(t) \tag{2.35b}$$

Taking the dimensionless frequency $x$:

$$x = \left(\frac{GM\omega}{c^3}\right)^{2/3} \tag{2.36}$$

Table 2.1: The data of the registered by LIGO events in 2015. UTC is the Coordinated Universal Time of the detection, ST – the sidereal time of the detection in hours, $\Delta_{\text{LH}}$ – the time delay of a signal arrival between antennas, $h$ – the strain (the amplitude of the signal), $f_0$ – the average frequency of the signal.

| Event | UTC | ST | $\Delta_{\text{LH}}$ [ms] | $h/10^{-21}$ | $f_0$ [Hz] |
|---|---|---|---|---|---|
| GW150914 | 09:50:45 | 3.33 | $6.9 \pm 0.5$ | 0.6 | 100 |
| LVT151012 | 09:54:43 | 5.24 | $-0.6 \pm 0.6$ | 0.3 | 100 |
| GW151226 | 03:38:53 | 3.89 | $1.1 \pm 0.3$ | 0.3 | 100 |

where $\omega = 2\pi f$ is the angular velocity of the rotation.

The time-dependent sinusoidal waveform is given by the angle $\psi(t)$ describing the mass motion on the circular orbit by the orbital phase, which depends on the time to the coalescence. The detailed formulas can be found i.e. (Blanchet (2014)).

For qualitative discussion, it can be noted that the detected strain with the known from the observations frequency and the mass estimation from the "chirp"-mass (see (2.43), Abbott et al. (2016d)) allows one to make conclusions about the distance to the objects under the assumption about the source as a coalescing binary. Some error may occur due to the unknown inclination of the orbit.

### 2.2.5 Analysis of GW events detected by LIGO in 2015

The basic relationships given in the chapter (2.2) allow us to make conclusions about the internal dynamics of the CBC by means of the data of the observed GW signal. In this part, I will consider how to apply the main relations presented above in order to get the information about the internal dynamics of a CBC analyzing the detected by LIGO signals.

The events GW150914, GW151226 and LVT151012 were detected by means of the interferometric antennas LIGO Livingston and Hanford (Abbott et al. (2016a)) in the period from September to December 2015. The observational data includes: the "strain" as the form of the signal (see eg Fig. 1 in Abbott et al. (2016a) for the GW150914), the frequency changing with the time, the time delay between the signal registrations at these antennas $\Delta_{\text{LH}}$ and the time of the event in UTC. The most important data for the calculations are summarized in Tab. (2.1). Assuming that the source of the GW event – the binary coalescence (CBC), the waveform can be divided into three stages: the "inspiral", "merger" and "ringdown" (see eg., Thorne (1989)). Here is considered the "inspiral" stage. Conditionally, can be fixed the time of the start of observations $t_0$ and the time until the merger stage – $t_c$. Using the data about the frequency change $f_{\text{GW}}$ with the time, there can be obtained the "chirp"-mass $\mathfrak{M}$ by (2.33) to make estimations of the sum mass $M$. For instance, for the event GW150914, there was concluded about the total mass as $M_{\text{total}} \approx 65 M_\odot$ indicating that the both incoming bodies should be relativistic compact objects (or RCO) or just "black holes". For the others events 2015, there were also made the conclusion about BH-BH binaries.

Then, with the estimated total mass $M$ of the binary and with the detected frequency $f$, the information about the distances to the objects can be obtained from the waveform (2.34). Some inaccuracy in the estimation of the distance is associated with the uncertainty in the inclination of the orbit and the orbital phase $\psi(t)$. Thus, the ordinal estimates of the distance under the assumption of the tensor GW from a CBC are reliable enough, and for the detected events 2015, the distances are defined to be $400 \div 1000$ Mpc.

Using the general relation of the signal amplitude (strain) $h$ to the distance to the source $r$ for tensor wave (2.15), the energy $E_{\text{GW}}[M_\odot c^2]$ of the GW can be estimated as:

$$h_0^{\text{tens}} \approx 0.54 \times 10^{-21} \left(\frac{\Delta E}{M_\odot c^2}\right)^{\frac{1}{2}} \left(\frac{0.1 \text{s}}{\tau}\right)^{\frac{1}{2}} \left(\frac{100 \text{Hz}}{f}\right) \left(\frac{400 \text{Mpc}}{r}\right) \quad (2.37)$$

where the normalizing parameters are chosen ones corresponding to the events LIGO 2015. The dependence "distance $r$ – strain $h$" is depicted on the Fig. (2.1), which shows that with the detected $h \approx 0.5 \times 10^{-21}$ and the estimated distances $400 \div 1000$ Mpc, the energy radiated into GWs corresponds to $1 \div 3 M_\odot c^2$ (where $M_\odot$ is the mass of the Sun).

Table 2.2: The physical parameters of CBC obtained by the analysis of the LIGO events GW150914, GW151226 and LVT151012

| Events | GW150914 | GW151226 | LVT151012 |
|---|---|---|---|
| Chirp-mass $\mathfrak{M}/M_\odot$ | 28.1 | 8.9 | 15.1 |
| Total mass $M/M_\odot$ | 65.3 | 21.8 | 37 |
| Primary mass $m_1/M_\odot$ | 36.2 | 14.2 | 23 |
| Secondary mass $m_2/M_\odot$ | 29.1 | 7.5 | 13 |
| Final mass $M_f/M_\odot$ | 62.3 | 20.8 | 35 |
| Luminosity distance $r/$ Mpc | 420 | 440 | 1000 |
| Source redshift $z$ | 0.09 | 0.09 | 0.20 |
| Radiated energy $E_{\rm GW}/M_\odot c^2$ | 3.0 | 1.0 | 1.5 |
| Peak luminosity $L_{\max}/$ erg/s | $3.6 \times 10^{56}$ | $3.3 \times 10^{56}$ | $3.1 \times 10^{56}$ |

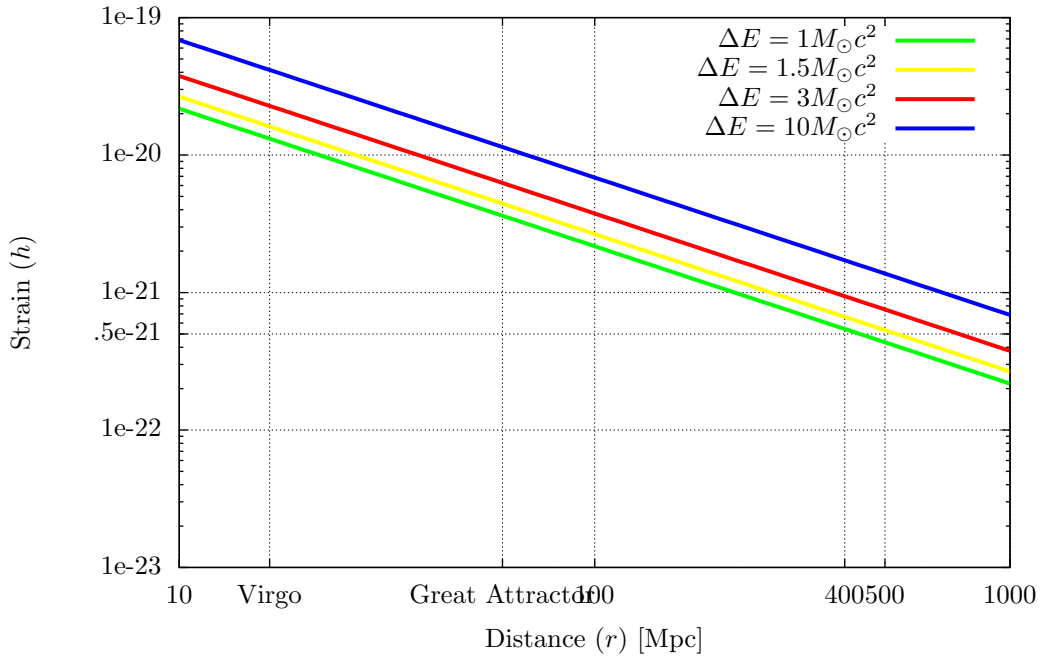

Figure 2.1: Relationship $h - r$ for tensor GW radiation from CBCs (2.39) at the several energies corresponding to the LIGO events 2015.

## 2.3   GW from Core-Collapse Supernova

Discussing GW radiation from supernova bursts, it is necessary to classify such objects concerning to the explosion mechanism, which can be symmetric or asymmetric. According to gravitation theory, the tensor wave can be radiated only as a result of an asymmetric collapse (see eg., Misner et al. (1973), Hawking and Israel (1989)), whereas, under the FGT and some tensor-scalar metric theories, there is predicted the existence of the scalar wave in the case of a spherically symmetric core-collapse Supernova (hereafter CCSN).

Examining the possible nature of the sources of the LIGO events, there will be analyzed such a prediction of the existence of the scalar wave from the CCSN. Firstly, it is motivated by the possibility of the sinusoidal signal similar to the detected ones, due to spherically symmetric core pulsations. While the tensor waveform by asymmetric collapse is expected to be more complex (see eg., Hawking and Israel (1989), Thorne (1989)). Secondly, as will be shown, the radiated energy as a result of the CCSN is estimated to be orders of magnitude more than ones due to the asymmetric collapse. Consequently, the GWs from CCSN is possible to detect by the antennas of the current sensitivity ($h \sim 10^{-23}$, LIGO and Virgo) at the distances up to 100 Mpc.

### 2.3.1 Energy and amplitude estimations for CCSN

According to the theoretical predictions, the tensor GW can be radiated only due to an asymmetric or axisymmetric core-collapse supernova (see eg., Hawking and Israel (1989), Thorne (1989)). However, despite the long-term theoretical study of the gravitational core-collapse of the stars, there is still no reliable estimates of the rate of asymmetry in such processes. Which in turn causes uncertainty in the estimates of the radiated energy in the form of GWs.

Thus, studying the dynamics of the asymmetric collapse of the rotating SN core, the authors, Zwerger and Mueller (1997), came to the conclusion that the energy released into the GW is around $E_{\text{GW}} = 10^{-11} \div 10^{-8} M_\odot c^2$, which corresponds to the amplitude of the tensor wave $4 \cdot 10^{-25} \leq h \leq 4 \cdot 10^{-23}$ from the source at the distance 10 Mpc. To a similar result came Bonazzola et al. (1993) for the case of an axisymmetric rotating core with the asymmetry rate $s < 0.1$. On the other hand, examining a fast-rotating core-collapse, Stark and Piran (1985) have given an estimate of the energy radiated in the "+" or "×" polarization mode as $E_{\text{GW}} \leq 10^{-3} M_\odot c^2$.

Besides this, there has been proposed the mechanism of the massive SN explosion by Imshennik and Nadezin (Imshennik (2010)), when as a result of the rapid rotation of the collapsing core, it is becoming asymmetric and may form a dipole configuration similar to the compact binary. Such a system might produce strong tensor radiation with a waveform similar to the case of coalescing neutron binary (NS-NS). Then, after the collapsing of such a system in the final RCO (relativistic compact object), there may occur pulsations giving scalar GW.

Scalar-tensor metric theories predict apart from tensor waves, the presence of the scalar radiation, which may arise as a result of the spherically-symmetric Core Collapse Supernova (CCSN) (Novak and Ibanez (2000)). In this case, the GW energy is expected to be up to $E_{\text{GW}} \leq 10^{-3} M_\odot c^2$.

In this way, despite the uncertainty in the explosion mechanism itself, the estimations of the energy emitted in scalar GWs as a result of a spherically-symmetric CCSN are on average by an order of the magnitude higher than such estimations for tensor GWs by an asymmetric or an axisymmetric rotating collapse. In both cases, the duration of a pulsation is estimated to be of the order of $0.5 \div 5$ ms, the duration of the whole pulse – 1 ms, and the GW frequency is around $f \approx 10^2 \div 10^3$ Hz (Zwerger and Mueller (1997)).

The general formulas for the amplitude estimation for the tensor (2.15) and the scalar (2.19) radiation show the expected difference in the magnitude of the detected strain. A typical GW signal from a CCSN pulsation can be represented as a unit pulse with the amplitude $h_0$, the frequency $f_0$ and the total duration $\tau$. There can be estimated the characteristic amplitude of the scalar GW (2.19) from a typical CCSN burst at the distances at around $\sim 1$ Mpc, with the duration $\tau = 0.1$ s and emitted during this time the energy $\Delta E_{\text{GW}} = 10^{-3} M_\odot c^2$ at the frequency $f = 100$ Hz:

$$h_0^{\text{sc}} \approx 1.36 \times 10^{-20} \left(\frac{\Delta E}{10^{-3}}\right)^{\frac{1}{2}} \left(\frac{0.1\text{s}}{\tau}\right)^{\frac{1}{2}} \left(\frac{100\text{Hz}}{f}\right) \left(\frac{1\text{Mpc}}{r}\right) \tag{2.38}$$

While the expected strain of the tensor GW (from an asymmetric CCSN) (2.15) will be approximately 2 times less (see eg., Schutz and Ricci (2010)):

$$h_0^{\text{tens}} \approx 6 \times 10^{-21} \left(\frac{\Delta E}{10^{-3}}\right)^{\frac{1}{2}} \left(\frac{0.1\text{s}}{\tau}\right)^{\frac{1}{2}} \left(\frac{100\text{Hz}}{f}\right) \left(\frac{1\text{Mpc}}{r}\right) \tag{2.39}$$

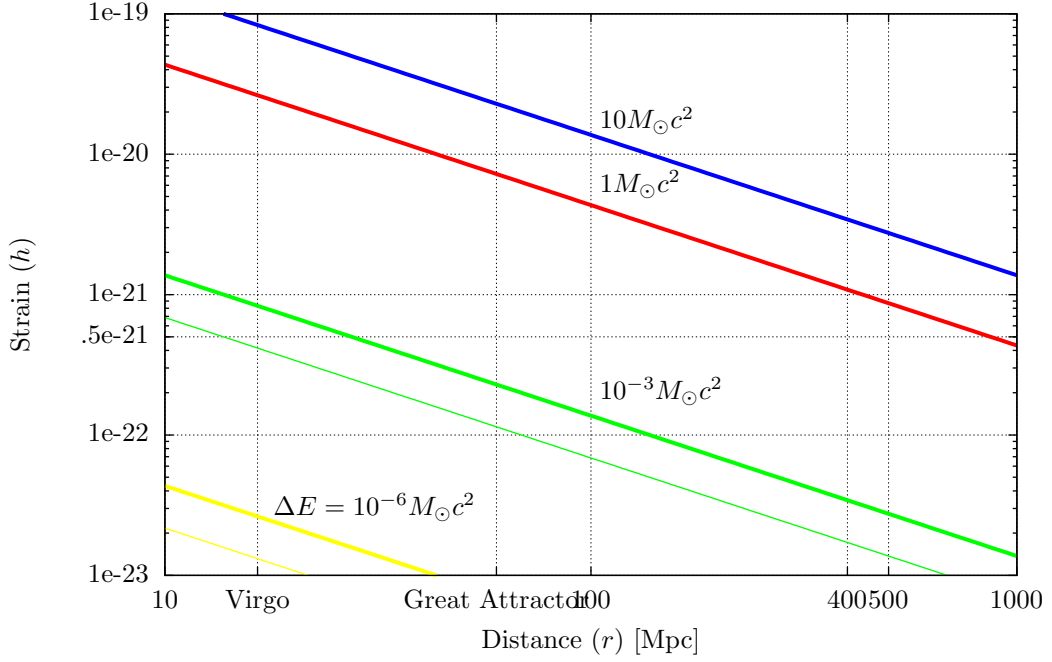

Figure 2.2: Relationship $h - r$ for the tensor, indicated by the thick lines, f. (2.39) and the scalar – by the thin lines, f. (2.38), GW radiation from CCSN. The lines of the same colour represent the same energy emitted in GWs.

Thus, with the considered parameters, the emitted GW of both kinds of polarization give the strain large enough to be detected by the modern interferometric antennas LIGO, Virgo with the current sensitivity threshold ($h \approx 10^{-23}$, Abbott et al. (2016a)).

The Fig. (2.2) depicts the relation "strain $h$ – distance $r$" in the logarithmic scale, where the colour lines indicate the emitted GW energy at the different order from $\Delta E_{\rm GW} = 10^{-6} M_\odot c^2$ to $\Delta E_{\rm GW} = 10^{-3} M_\odot c^2$ according to the given above estimations in the frames of GRT and the scalar-tensor metric theories. The thick lines represent the relation for the scalar radiation, the thin ones of the same colour – for the tensor at the same GW energy. Additionally, there are marked roughly the distances, at which the local clusters of galaxies are located: the Virgo cluster (16 Mpc) and the Great Attractor (47 ÷ 74 Mpc). Interestingly, for these possible energies of GWs, the distances to the objects are at up to $r \leq 26$ Mpc, at the observed in 2015 strain $h \approx 0.5 \cdot 10^{-21}$.

In contrast to the most common view regarding the mechanism of the CCSN in the frames of GRT and the scalar-tensor metric theories, in the FGT approach (Baryshev (2017)), the upper limit on the radiated in GW energy is established only by the mass of the object itself, which may amount to the several solar masses.

It can be shown that the characteristic amplitude of the scalar wave from a source at the distance of $\sim 100$ Mpc with the GW energy of the order of several solar masses is:

$$h_0^{\rm sc} \approx 4.34 \times 10^{-21} \left(\frac{\Delta E}{M_\odot c^2}\right)^{\frac{1}{2}} \left(\frac{0.1 {\rm s}}{\tau}\right)^{\frac{1}{2}} \left(\frac{100 {\rm Hz}}{f}\right) \left(\frac{100 {\rm Mpc}}{r}\right) \qquad (2.40)$$

The corresponding relation $h - r$ for the GW energy $\Delta E_{\rm GW} = 1 \div 10 M_\odot c^2$ is represented in Fig. (2.2). Thus, the GW from a CCSN at the distance $r = 100$ Mpc radiating GW with the energy $\Delta E_{\rm GW} \sim 10^{-2} M_\odot c^2$ is expected to give the detected strain $h \approx 0.5 \cdot 10^{-21}$.

Comparing the results for the scalar-tensor radiation from CCSN in the frames of both the metric gravitation theories and the FGT, the following conclusion can be made. In contradistinction to the tensor-scalar metric theories, there is no absolute restriction on the radiated into GW energy in the FGT. This allows one to consider the objects as GW sources at farther distances and, consequently, with larger total masses, which might provide corresponding energy on the GW radiation. That in turn establishes the lower limit on the rest mass of the SN collapsing core. Identifying the GW signal as the CCSN, for instance, by the analysis of the follow-up events in the electromagnetic spectrum, the relationship (2.40) suggests a test for the existence of such objects as supermassive SN. Thus, with the known distance to the object from the EM observations of the transient, and with the detected GW amplitude, it is possible to estimate the energy radiated into GW in the units of solar masses.

Further in this work, there will be given the analysis of the LIGO events in 2015 to get estimates on the possible physical parameters of such a CCSN.

### 2.3.2 Scalar wave from CCSN in the FGT

As has been discussed, as a result of a spherically-symmetric core-collapse SN, there is predicted the scalar GW radiation (see review Baryshev (2017)). In this case, the expected signal might be in the form of a sinusoidal pulse with the increasing frequency of the pulsations. Further discussion is motivated by the LIGO observations in 2015, which is a fairly correct sinusoidal pulse with the known average frequency and amplitude. In this part will be considered the general relations between the detected values of a GW signal: its "strain" $h$, (average) period $P_0$, average frequency $f_0$, and the physical parameters of the pulsating object: its density $P_0$, radius $R_0$, as well as the distance $r$ to the object.

For a CCSN with the characteristic period of the pulsations $P_0 \sim 1/f_0 \sim 1/\sqrt{G\rho_{\text{eff}}}$ (Baryshev, Yu. V. (1990)), there can be determined the effective density $\rho_{\text{eff}}$ taking into account the inhomogeneity of the mass distribution along the radius of the object:

$$\rho_{\text{eff}} \sim \frac{1}{P_0^2 G} \tag{2.41}$$

Let us introduce the parameter $\gamma$ characterising the relationship between the effective density $\rho_{\text{eff}}$ and the average density $\rho_0 = M_0/(\frac{4}{3}\pi R_0^3)$, where $R_0$ is the radius of the object, and $M_0$ – its (average) mass.

$$\gamma = \frac{\rho_{\text{eff}}}{\rho_0} \tag{2.42}$$

The next parameters can be introduced: $\alpha$ characterising the pulsations velocity $v_0 \sim R_0/P_0$ relative to the speed of light $c$, and $\beta$ – for the ratio of the average radius $R_0$ of the object to its gravitational radius $R_G$:

$$\alpha = \frac{v_0}{c} \tag{2.43a}$$

$$\beta = \frac{R_0}{R_G} \tag{2.43b}$$

The compatibility condition for the entered parameters can be written as:

$$\frac{\gamma}{\beta} = \frac{4}{3}\pi\alpha^2 \tag{2.44}$$

Which means that the relationship $\gamma/\beta$ can be determined by the known or estimated parameter $\alpha$. The parameter limiting conditions: $0 \leq \alpha \leq 1$; $0 \leq \gamma \leq 1$; $\beta \geq 1$ are presented in Fig. (2.3).

The amplitude of a GW at the distance $r$ from the source can be obtained as:

$$h_0 \sim \frac{R_G}{r}\alpha^2 \tag{2.45}$$

where $R_G = GM_0/c^2$ is the gravitational radius of the collapsing core.

Using the derived above relationships, there can be represented the relation "the distance to the object $r$ – the registered strain $h$" in the form depending only on the observed period of the pulsations $P_0$ and the parameter of the changing rate $\alpha$:

$$h_0 \sim \frac{4}{3}\pi c \cdot \frac{P_0}{r} \cdot \alpha^5 \sim c \cdot \frac{P_0}{r} \cdot \frac{\gamma}{\beta} \cdot \alpha^3 \tag{2.46}$$

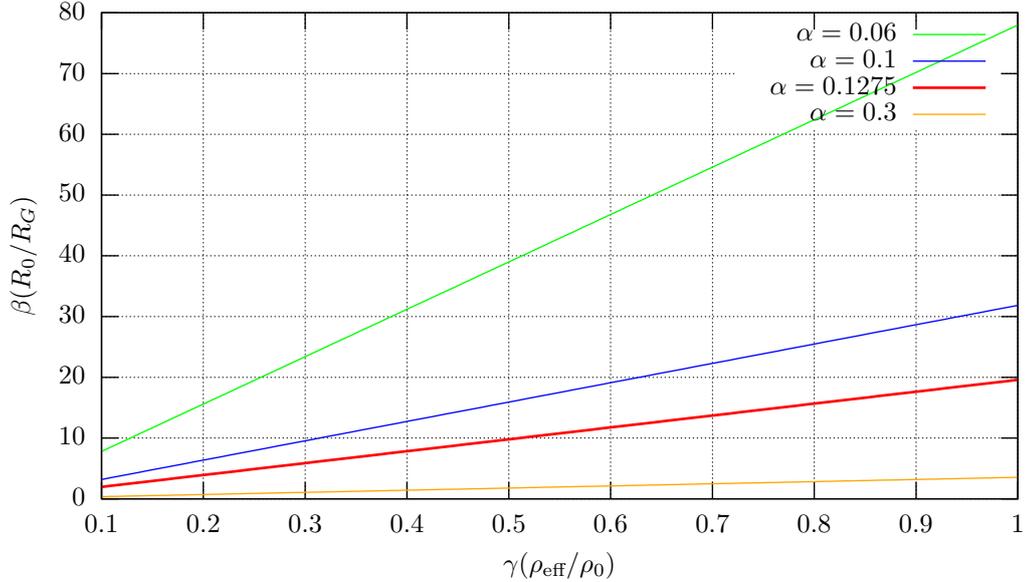

Figure 2.3: Illustration of the compatibility condition (2.44): dependence of the parameters $\beta, \gamma$ on the $\alpha = v_0/c$ for a CCSN.

Thereby, with the observed data $h, P_0$ and an supposed GW energy $\Delta E_{\mathrm{GW}}$, there can be estimated (2.46) the introduced source parameters $\alpha, \beta, \gamma$ connected by the compatibility condition (2.44). Which give the estimates of the radius $R_0$, mass $M_0$ and density $\rho_0$ of the CCSN.

### 2.3.3 Analysis of the GW events detected by LIGO in 2015

According to the considered above approaches to the study of a CCSN, metric and field, there is established a different limit on the radiated into GWs energy. Thus, according to the scalar-tensor metric theories, this limit is estimated to be $\Delta E_{\mathrm{GW}} \leq 10^{-3} M_\odot c^2$, while in the FGT, the amount of the radiated energy is limited only by the rest mass of a collapsing, which can amount several solar masses. In this connection, it can be shown what difference is expected in the parameters of a CCSN radiating GWs of different energy, with a frequency and an amplitude similar to the detected by LIGO in 2015.

The calculations have been made using formulae (2.38) for scalar and (2.39) for tensor GW mode, which illustrate the dependence of the GW amplitude $h_0$ on the distance to the object $r$ for a typical CCSN with radiated GW energy of the order $\Delta E_{\mathrm{GW}} = 10^{-3} M_\odot c^2$, according to the discussed above the estimates of the maximal energy possible to be radiated in GWs in the frames of the scalar-tensor metric theories. Besides this, there has been done the calculation for a scalar wave from spherically-symmetric CCSN with the radiation energy $\Delta E_{\mathrm{GW}} = 1 M_\odot c^2$, which is possible in the frame of the FGT.

The average data of the GW signals detected by LIGO in 2015–2017: frequency $f_0 = 100$ Hz, pulse duration $\tau = 0.1$ s (Tab. 2.1). The results of calculations are shown on Fig. (2.2). As has been mentioned, with the same detected strain, the GW with higher energy will come from a more distant object.

There can be estimated the parameters of a CCSN at such distances, radiating a scalar GW with an average period $P_0 = 1/f_0 = 0.01$ s. The relationship for a CCSN (2.46) give the parameter $\alpha = v_0/c$ as well as the parameter ratio $\gamma/\beta$ (2.44). The results for typical values of radiated energy: $\Delta E_{\mathrm{GW}} = 10^{-6}, 10^{-3}, 1 M_\odot c^2$ are given in the Tab. (2.3). The calculations have been made for the average detected strain value of $h = 0.6 \cdot 10^{-21}$ (GW150914), with the assumed $\gamma \equiv 1$, i.e. $\rho_0 = \rho_{\mathrm{eff}} = 0.15 \cdot 10^{12}$ g/cm$^3$.

There should be noted that these calculations are model and use average values of the detection parameters without taking into account the variation with the time. To sum up, a CCSN radiating GW with the energy of the order of the solar mass should have a high pulsation rate, a radius close to the gravitational one, and a mass close to the extreme estimates for a massive pre-star of the CCSN, to give a signal with the detected amplitude.

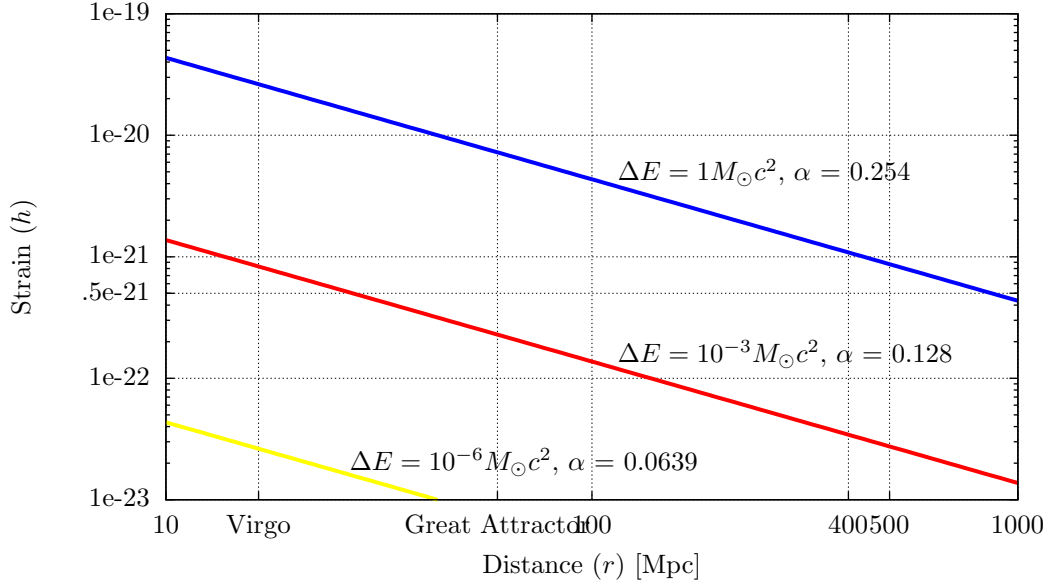

Figure 2.4: Illustration of the relationship (2.46) for the several cases of the radiated energy in the scalar GWs from a CCSN.

Table 2.3: Calculated parameters for a CCSN under the condition that the detected GW signal has average frequency $f = 100$ Hz. $r$ is the distance to the object corresponding to the detected strain $h = 0.6 \cdot 10^{-21}$ (GW150914). Mass of the Sun is $M_\odot \approx 2 \cdot 10^3$ g, the gravitational radius $R_G = GM_0/c^2$, and the assumed parameter $\gamma = \rho_{\text{eff}}/\rho_0 \equiv 1$ gives the effective frequency $\rho_{\text{eff}} = 0.15 \cdot 10^{12}$ g/cm$^3$.

| $\Delta E \, [M_\odot c^2]$ | $r$ [Mpc] | $\alpha = v_0/c$ | $\beta = R_0/R_G$ | $M_0/M_\odot$ |
|---|---|---|---|---|
| $10^{-6}$ | 0.72 | 0.064 | 58.436 | 2.220 |
| $10^{-3}$ | 22.88 | 0.128 | 14.679 | 17.635 |
| 1 | 723.57 | 0.254 | 3.687 | 140.078 |

## 2.4 Conclusion

Within the framework of the GR, there exists only tensor GW radiation, which can occur as a result of a compact binary coalescence (CBC) or an asymmetric core-collapse supernova (CCSN). Analysing the waveform of the LIGO signals in 2015, there has been made the conclusion about the nature of all these sources being CBCs. For such a system, with the registered signal parameters, there can be drawn sufficiently reliable conclusions about the size of the system, the masses of the incoming bodies, the distances to it, and the GW energies. Thus, in the case of the GW150914, the distance to the generating CBC is estimated to be $\sim 440$ Mpc. An important test of the model of a CBC is the identification of a detected GW event with the optical and X-ray transients, which are possible only in the case of the coalescence of RCOs without the events horizon, which is possible within the frame of the field theory of gravity ("gravidynamics").

In the frame of the GR modifications (the scalar-tensor metric theories), as well as in the field approach to describing gravity (the FGT), there is predicted the existence of scalar GW radiation from a spherically-symmetric pulsating core (CCSN), with the waveform expected to be close to a sinusoidal with a varying frequency similar to the waveform from a CBC. The principal difference between the predictions of the metric theories based on the GR and the FGT is the limit for the radiated in GWs energy. According to the scalar-tensor metric theories, there is presupposed the GW energy radiated from a CCSN to have a limit of $\sim 10^{-3} M_\odot c^2$, while the FGT makes the limitation on the radiated energy only by the rest mass of the collapsing SN core itself.

The carried out in this paper evaluations for the scenarios of scalar wave radiation as a result of a CCSN from the point of view of the metric theories and in the FGT approach showed that at the same recorded wave amplitude but at different limiting energies, the sources should be at different distances and have different internal characteristics such as mass, radius and kinetic energy of pulsations. The observational test of the existence of spherically-symmetric pulsations in the core of supermassive SNs will be the identification of the GW event with the associated SN counterpart in the electromagnetic spectrum.This also will allow us to estimate the distance to the object and, as a consequence, the limits of the mass and density of the object.

# Chapter 3

# Method for source localization by GW polarization state

The purpose of the method is to select a GW source localization on the sky in the case of detection by two interferometric antennas, when it is only possible to make suggestions about an allocation of the source along a certain apparent circle rather than in one point. This apparent circle (hereafter AC) is determined by the time delay $\Delta$ between signal registration at the two antennas and sidereal time (ST) of the event. Further contraction of the area is carried out with an assumption about polarization state of the incoming GW together with the strain ratio detected at these antennas. When detected GW event is identified with electromagnetic "follow-up", then the source location can be uniquely determined. That allows us to make more reliable conclusions about the nature of the source as well as the polarization state of the detected GW.

## 3.1 GW detection

Generally, in metric gravitation theories, there are six polarization states (Eardley et al. (1973a), Will (2014)). Three of them are transverse to the direction of wave propagation, with two representing quadrupolar deformations (tensor transverse wave) and one representing a monopolar "breathing" deformation (scalar transverse wave). Other three modes are longitudinal, including the stretching mode in the propagation direction (scalar longitudinal wave). In the frame of GRT, there are only 2 tensor transverse polarization states, + ("plus") and × ("cross"), under consideration (Misner et al. (1973)), while Feynman's field theory (FTG, see (2.1.2)) as well as some modified scalar-tensor metric theories predict the existence of scalar transverse and/or longitudinal modes. In this work, I will focus on the possibility to disentangle between four polarization states: tensor "plus"- (+) and "cross"- (×), scalar longitudinal and transverse (see Fig. (3.1)).

Consider the information received from the GW signal detected by a Michelson-type interferometric antenna. The orthogonal arms with four test masses at their ends are pointing in X- and Y-direction in a Cartesian coordinate system, which is at the rest in the local proper reference frame of the detector, Fig. (3.2)). The GW passing through the antenna displaces the test masses, thereby changing the length of each arm from its initial length $L_0$ (for LIGO detectors $L_0 = 4$ km). The monitored by laser difference between lengths of these arms $\Delta L(t) = L_X - L_Y$ gives the observed at the antenna strain $h(t) = \Delta L(t)/L_0$. The position of a source S relative to the detector is shown schematically in Fig. (3.2), where $\zeta$ is the zenith angle, $\Phi$ – the azimuth of the source in horizontal coordinate system of the antenna.

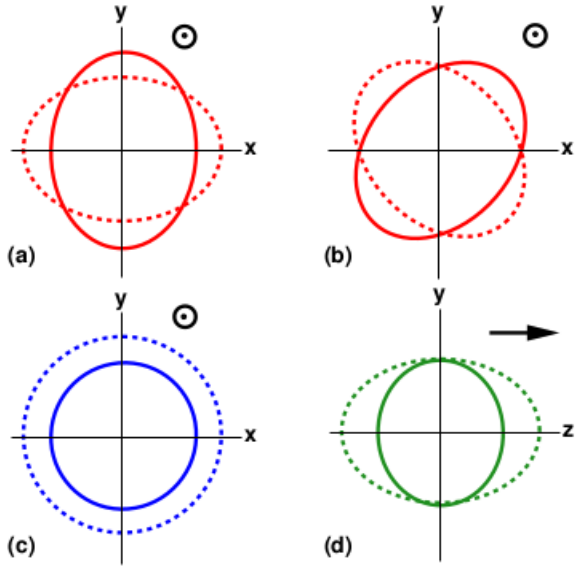
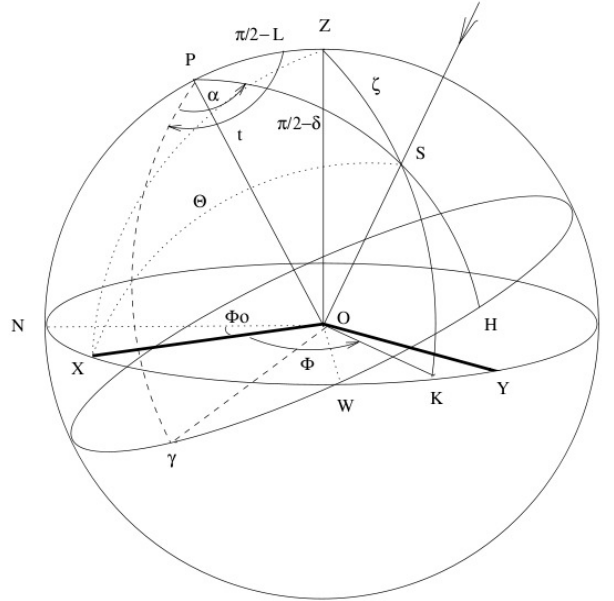

Figure 3.1: The four (from possible six) polarization modes of GW. There is shown the displacement that each mode induces on a ring of freely moving test bodies. The wave is propagating in the +z direction: in (a), (b), and (c), – out of the drawing plane; in (d) – in the plane. In GRT, only (a) – tensor "+" and (b) – tensor "×" are presented, whereas some massless scalar-tensor gravitation theories predict the existence also (c) – scalar transverse and/or (d) – scalar longitudinal modes (Will (2014)), (Eardley et al. (1973b)).

Figure 3.2: Equatorial and horizontal coordinate systems of an interferometric antenna for the GW source S. $Z$ is the zenith, $P$ – the northern pole, $\gamma$ defines the sidereal time, $\alpha$ – the right ascension (RA), $\delta$ – the declination (DEC), $\zeta$ – the zenith angle. The reference direction of the detector is the direction OX with the azimuth $\Phi_0$.

### 3.1.1 Antenna-pattern functions for 2-arm interferometric detector

For laser interferometers, the general form of the signal $\bar{h}$ is a composition of antenna-pattern functions $F$ with waveform $h(t)$ for the corresponding polarization mode (Will (2014)):

$$\bar{h}(t;\zeta,\Phi,\Psi) = F_{\text{SL}}(\zeta,\Phi)h_S(t) + F_{\text{ST}}(\zeta,\Phi)h_S(t) + F_+(\zeta,\Phi,\Psi)h_+(t) + F_\times(\zeta,\Phi,\Psi)h_\times(t) \quad (3.1)$$

Antenna-pattern functions $F(\zeta,\Phi,\Psi)$ are determined for each polarization state (see, e.g., Will (2014)) by the angles $(\zeta,\Phi)$ of a source position in horizontal coordinate system of the antenna, Fig. (3.2)), and polarization angle $\Psi$ in the case of tensor GW (for definition see p. 366 Hawking and Israel (1989)):

$$F_{\text{SL}}(\zeta,\Phi) = \frac{1}{2}\sin^2\zeta\cos 2\Phi \quad (3.2a)$$

$$F_{\text{ST}}(\zeta,\Phi) = -\frac{1}{2}\sin^2\zeta\cos 2\Phi \quad (3.2b)$$

$$F_+(\zeta,\Phi,\Psi) = \frac{1}{2}(1+\cos^2\zeta)\cos 2\Phi\cos 2\Psi - \cos\zeta\sin 2\Phi\sin 2\Psi \quad (3.2c)$$

$$F_\times(\zeta,\Phi,\Psi) = \frac{1}{2}(1+\cos^2\zeta)\cos 2\Phi\sin 2\Psi + \cos\zeta\sin 2\Phi\cos 2\Psi \quad (3.2d)$$

where SL – "scalar longitudinal", ST, "scalar transverse".

Antenna-pattern functions are illustrated in Fig. (3.3). The beam patterns for tensor "+" and "×" are different, that reflects the fact that it is possible to distinguish tensor polarization modes by means of typical two-arms interferometric antenna. On the other hand, antenna-pattern functions for scalar waves differ only by sign (3.2a, 3.2b), as a result, beam patterns for longitudinal and transverse modes are the same, Fig. (3.3c). To tackle this problem, there will be considered the case of an one-arm interferometric antenna, which could be realized.

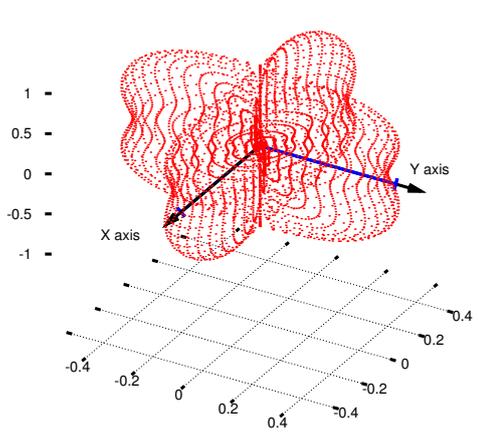

(a) tensor + wave

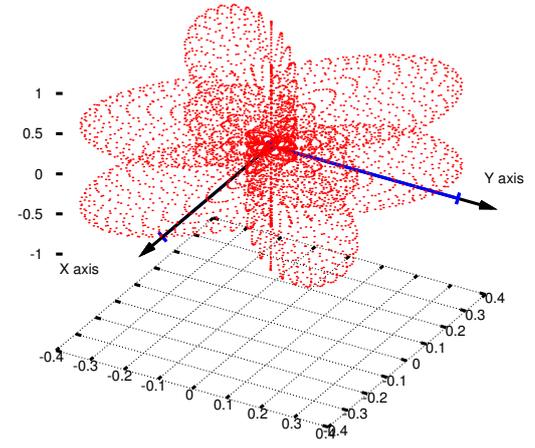

(b) tensor × wave

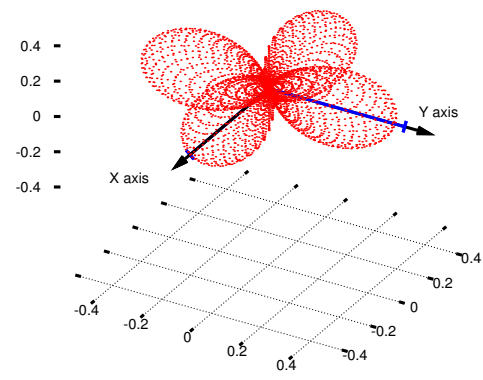

(c) scalar wave for 2-arm antenna

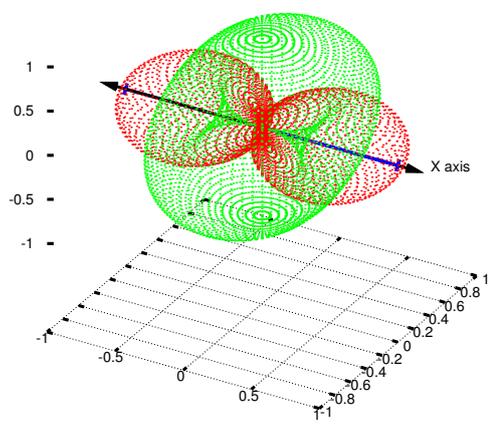

(d) scalar transverse (green) and longitudinal (red) waves for 1-arm antenna

Figure 3.3: Antenna patterns for different polarizations of an incoming GW. Blue lines indicate the arms of the detectors along the X- and Y-axis. Red points – antenna response factors depending on the location of the GW source on the sky.

### 3.1.2 Antenna-pattern functions for 1-arm interferometric detector

To solve the problem of indiscernibility between scalar longitudinal and transverse modes by a two-arms interferometer, there will be proposed a modification of an interferometric detector as one-arm antenna (one-arm mode) having one working arm with two test masses. Then the observed strain is given by the length change of the working arm (X-axis) relative to the length $L_0$ of the fixed (former Y-axis) arm: $\Delta L(t) = L_X - L_0$. The amplitude of the arm-length variation $h^0 = \Delta L_{max}/L_0$ can be used as a normalization constant.

Antenna-pattern functions on scalar GW for 1-arm interferometric detector (Baryshev and Paturel (2001)):

$$F_{\text{SL}}(\zeta, \Phi) = \cos\Theta = \sin\zeta \cos\Phi \tag{3.3a}$$

$$F_{\text{ST}}(\zeta, \Phi) = \sin\Theta \tag{3.3b}$$

where $\Theta$ – the angle between the direction of a GW propagation and X-axis of the antenna, Fig. (3.2)).

Fig. (3.3d) shows that the beam patterns are different for scalar transverse and longitudinal waves. Therefore, the application of one-arm interferometer to GW observations would allow us to disentangle these scalar modes.

## 3.2 Method description

The method uses the information about the strain $h$ of the detected GW signal at each antenna in the network, sidereal time (ST) of the arrival, time delay $\Delta$ between signal registration at the antennas, as well as antenna position in the equatorial coordinate system.

The detected time delay $\Delta$ between registrations determines a radius of an apparent circle (hereafter AC) on the unit sky sphere, along which the source of GW might be located. The centre of the AC is defined by the direction of the vector joining the two antennas at the sidereal time (ST) of the event. In an equatorial coordinate system, each point at the AC is defined by right ascension (RA) $\alpha$ and declination (DEC) $\delta$. Regarding the detector, the considered point as possible source S has horizontal coordinates: zenith angle $\zeta$ and azimuth $\Phi$, Fig. (3.2)). Thereby, beam patterns depending on the antenna-pattern functions $F(\zeta, \Phi, \Psi)$ are different for distinct antennas (ch. 3.1).

According to the principle of an antenna-interferometer (ch. 3.1), there is detected strain $h(t)$, which can be decomposed into time-dependent part – the normalized waveform $s(t)$, and time-independent – the geometric factor (or G-factor) $G(\zeta, \Phi, \Psi)$ characterizing the position of the source on the sky.

$$h(t) = \frac{\Delta L(t)}{L_0} = h^0\, s(t)\, G(\zeta, \Phi, \Psi) \tag{3.4}$$

where $h^0$ is the amplitude of the signal.

The geometric factor $G(\zeta, \Phi, \Psi)$ is determined by the relative orientation of an antenna with respect to the position of the source on the sky at the sidereal time (ST) of the detection, angles $(\zeta, \Phi)$, as well as by the polarization angle $\Psi$ for tensor GW. In the particular case of an incoming GW in a single polarization mode, the G-factor is equivalent to the antenna-pattern function ($G \equiv F$) for this mode. In a general case, the G-factor represents a composition of antenna-pattern functions $F$ weighted by coefficients identifying the entering polarization states (3.1).

Regarding the time-dependent part of a strain, it is worth to mention that the normalized waveform $s(t)$ depends only on the nature of the source and, consequently, is the same at each antenna in the network for a particular GW event.

Table 3.1: Parameters of the LIGO and Virgo GW antennas

| Name | Latitude | Longitude; Azimuth |
|---|---|---|
| LIGO L1 | 46°27′19″N | 119°24′28″W; N36°W |
| LIGO H1 | 30°33′46″N | 90°46′27″W; W18°S |
| Virgo | 43°37′53″N | 10°30′16″E; N19°E |

Table 3.2: Detection parameters of LIGO events. ST is the sidereal time of the event, $h^0$ – the strain as the maximal amplitude normalized by $10^{-21}$, $\Delta_{\text{LH}}$ – the time delay between registrations at Livingston and Hanford antennas.

| GW event | UTC | ST [hrs] | $\Delta_{\text{LH}}$ [ms] | $h^0$ |
|---|---|---|---|---|
| GW150914 | 09:50:45 | 3.3315 | $6.9^{+0.5}_{-0.4}$ | 0.6 |
| LVT151012 | 09:54:43 | 5.2377 | $-0.6 \pm 0.6$ | 0.3 |
| GW151226 | 03:38:53 | 3.8851 | $1.1 \pm 0.3$ | 0.3 |

Then the following relation holds at the fixed sidereal time:

$$\frac{h_1}{h_2} = \frac{G_1}{G_2} = \frac{G(\zeta_1, \Phi_1, \Psi_1)}{G(\zeta_2, \Phi_2, \Psi_2)} \tag{3.5}$$

where 1 and 2 indicate the values related to the considered antennas.

Thus, the calculated ratio of $G_1/G_2$ for a certain point on an AC predicts the observed strain ratio $h_1/h_2$. Therefore, it is possible to highlight such points on the AC, where the calculated $G_1/G_2$ approximately is equal to the observed $h_1/h_2$. This is the essence of the method. It is important to note that the detected strain ratio does not depend on the nature of the source, but only on the antenna position relative to the source and polarization state of the incoming GW.

## 3.3 Application of the method to the LIGO events

In this chapter, the proposed method will be applied to the GW events detected by LIGO in 2015.

### 3.3.1 The used data

Consider three GW events 2015: GW150914, GW151226 and LVT151012 (Abbott et al. (2016a)). The used in the method data are given in the Tab. (3.2). For the reported GW events, there were only two antennas, LIGO Hanford and Livingston, in operation, the parameters of their position on the Earth are given in the Tab. (3.1). The time delay $\Delta_{\text{LH}}$ between the signal registration at these antennas together with the sidereal time of the event provide the AC in the sky, along which the GW source will be searched. In Fig. (3.4) we present the ACs for the events GW150914, LVT151012, and GW151226 in the first equatorial coordinate system with respect to the Earth (hour angle $t$ – declination $\delta$), as it is also shown by the LIGO-Virgo Scientific Collaborations (Abbott et al. (2016c)). Antenna positions are clearly indicated by the red triangles, H1 (Hanford) and L1 (Livingston), the points H1-L1 and L1-H1 mark the poles of the line joining these two detectors (in other words, the points of the maximal time delay).

The same ACs are also shown in the most common view – in the equatorial coordinate system using right ascension (RA) $\alpha$ and declination (DEC) $\delta$, Fig. (3.5), which is consistent with the general results by LIGO (Abbott et al. (2016a), Fig. 5).

The next step in the method application is the calculation geometrical factors for each point along the AC regarding the assumed polarization state of the incoming GW. There will be taken into consideration the case of "pure" polarization mode such as tensor "plus" or scalar. Additionally, there will be considered a more complicated case of a tensor mode mixture, which is expected by a coalescing binary.

The final step is to compute the ratio $G_L/G_H$ for each point of the AC and select such points, where this ratio is approximately equal to the observed $h_L/h_H \approx 1.0 \pm 10\%$ for all three events.

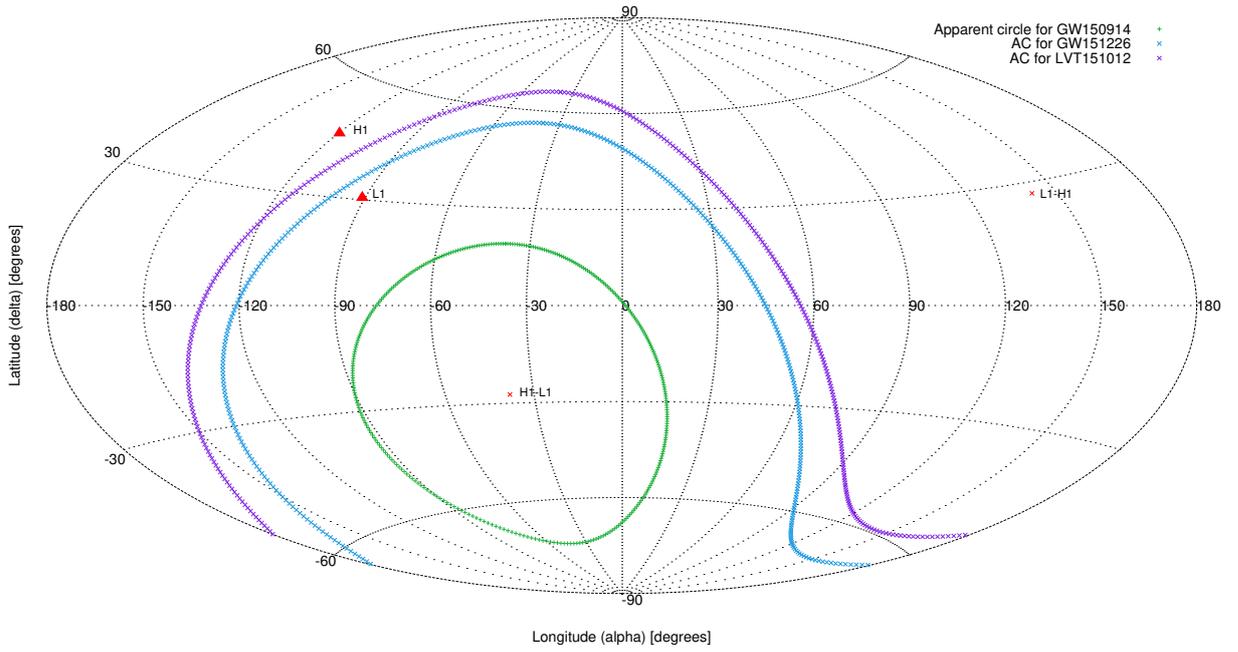

Figure 3.4: ACs of GW150914, LVT151012 and GW151226 shown in Aitoff projection with respect to the Earth at the time of detection. Red triangles, H1 and L1, indicate the positions of the LIGO LI (Livingston) and H1 (Hanford) interferometers, H1-L1 and L1-H1 mark the poles of the line joining these two detectors (the points of the maximal time delay).

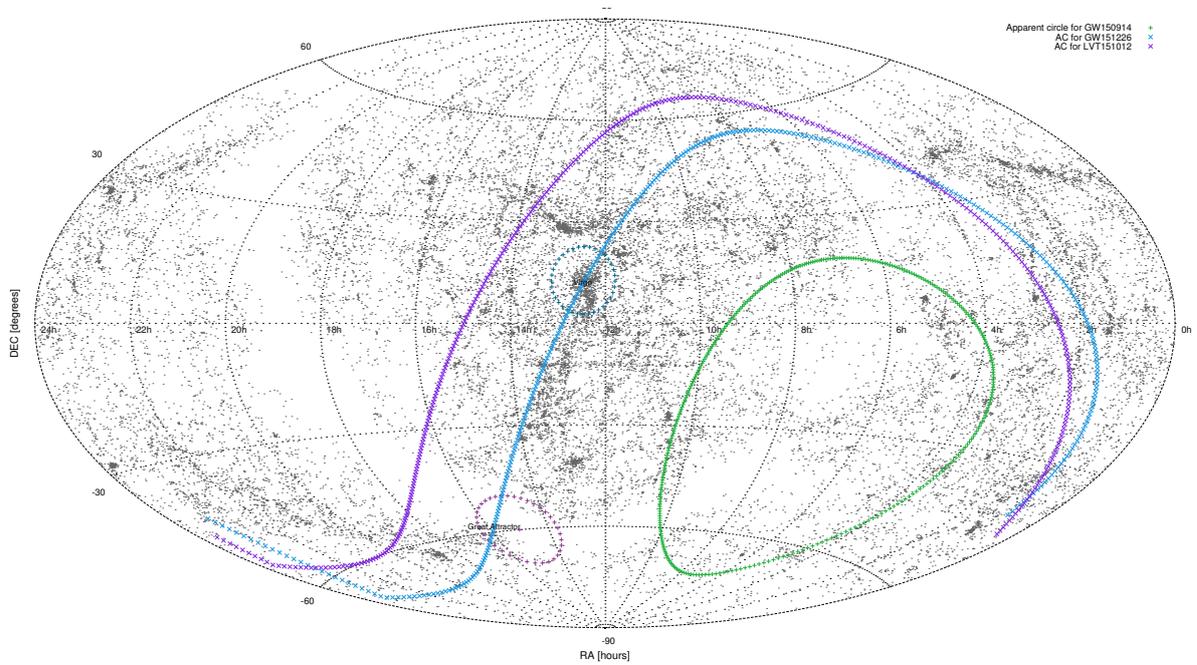

Figure 3.5: ACs of the allowed source positions for LIGO events: GW150914, LVT151012 and GW151226 in equatorial coordinates (Aitoff projection) with the projections of the galaxies from the 2MASS catalogue at the distance up to 100 Mpc ($z \leq 0.025$).

### 3.3.2 Calculations for scalar polarization

As has been mentioned in Ch. (3.1), a typical two-arms interferometric antenna does not distinguish between scalar transverse and longitudinal states, therefore, the results of the calculation are valid for these both modes.

According to the (3.4), G-factor of "pure" scalar polarization mode is equal to the antenna-pattern function (3.2a) for two-arms interferometric antenna: $F_{\text{sc}} \equiv F_{\text{SL}} \equiv F_{\text{ST}}$. Then the registered strain will be of the form:

$$h(t)_{\text{sc}} = h^0_{\text{sc}}\, s(t)_{\text{sc}}\, F_{\text{sc}}(\zeta, \Phi, \Psi) \tag{3.6}$$

For an one-arm interferometric detector (see 3.1.2), beam patterns are different for scalar transverse and longitudinal modes, and the calculations are hold using (3.3a) and (3.3b) respectively.

At the fixed sidereal time $t$, the main relation for scalar polarization wave is:

$$\frac{h_1}{h_2} = \frac{F_{\text{sc}}(\zeta_1, \Phi_1)}{F_{\text{sc}}(\zeta_2, \Phi_2)} \tag{3.7}$$

The results for the LIGO events in 2015 are presented in Fig. (3.6b) in equatorial coordinates. The red points highlight the localization area of a GW source of the event in the assumption about scalar longitudinal or transverse polarization of the incoming GW.

### 3.3.3 Calculations for tensor "plus" or "cross" polarization

Generally, the strain of tensor transverse GW is the combination of two polarization modes: "plus" $h_+$ and "cross" $h_\times$ with their own antenna response functions $F_+$ and $F_\times$. Therefore, the form of a detected strain $h(t)$ strongly depends on the waveform of each polarization mode, i.e. on the nature of the source. Further, there will be considered the case of a complex tensor wave from a compact binary (3.3.4). Another kind of objects considered to radiate a tensor wave – an asymmetric collapse of a supernova core (see Hawking and Israel (1989), Misner et al. (1973)) – is difficult to be analyzed and, consequently, to predict the accurate waveform. However, as a model task, there can be considered a "pure" tensor polarization mode: whether "plus" or "cross", assuming that there might be the case when this tensor mode dominates in an outcoming wave from such an object.

Then, similarly, as in the scalar case, the G-factors ratio for a single tensor mode depends only on the corresponding antenna-pattern function $F_{+,\times}$. At the known sidereal time of the detection, the main ratio for the method application:

$$\frac{h_1}{h_2} = \frac{F_{+,\times}(\zeta_1, \Phi_1, \Psi_1)}{F_{+,\times}(\zeta_2, \Phi_2, \Psi_2)} \tag{3.8}$$

The results of calculations for the tensor "+" mode for the LIGO events in 2015 are shown in Fig. (3.6a).

### 3.3.4 Calculations for mixed tensor wave from binary coalescence

In this section, there will be analysed the possibility of method application to the case of a complicated tensor radiation from a coalescence binary.

According to the (3.1), the strain of a tensor GW is a combination of "plus" $h_+$ and "cross" $h_\times$ together with antenna-pattern functions, $F_+$ and $F_\times$, in the proper reference frame of the detector:

$$h(t) = h_+(t)F_+(\zeta, \Phi, \Psi) + h_\times(t)F_\times(\zeta, \Phi, \Psi) \tag{3.9}$$

where $h_+, h_\times$ are calculated by (2.34) for each polarisation mode in a chosen post-Newtonian approximation, and $F_+, F_\times$ – by (3.2c, 3.2d).

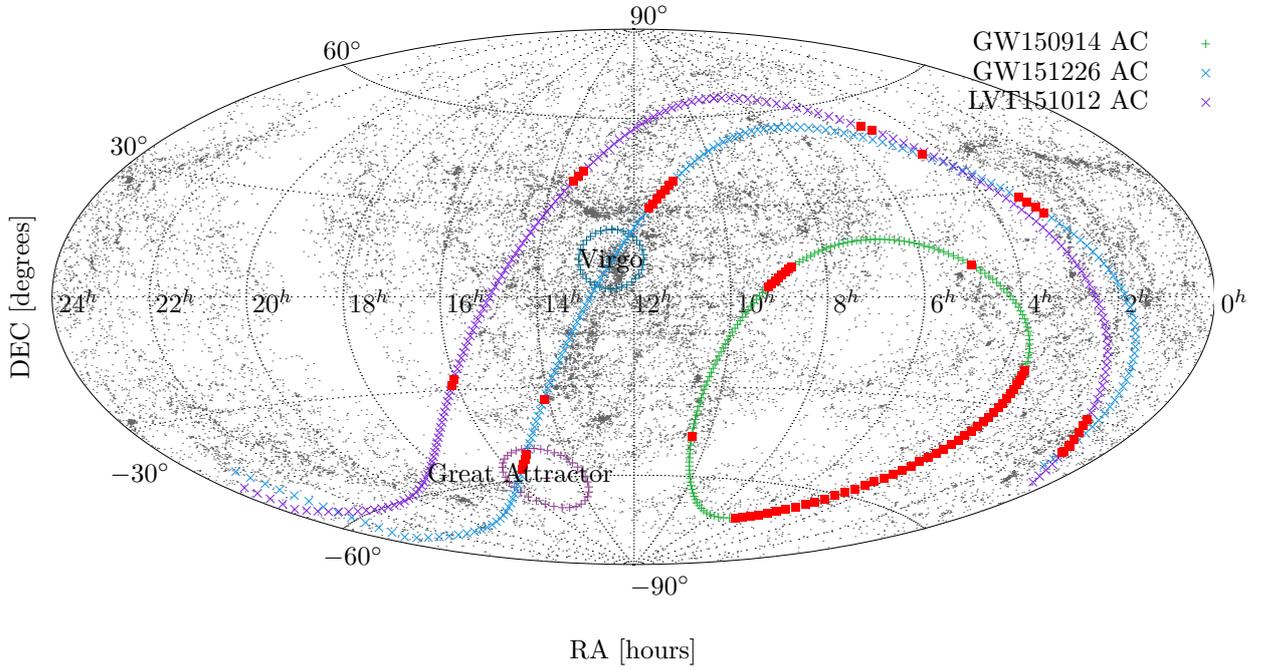

(a) for the case of a tensor "+" GW

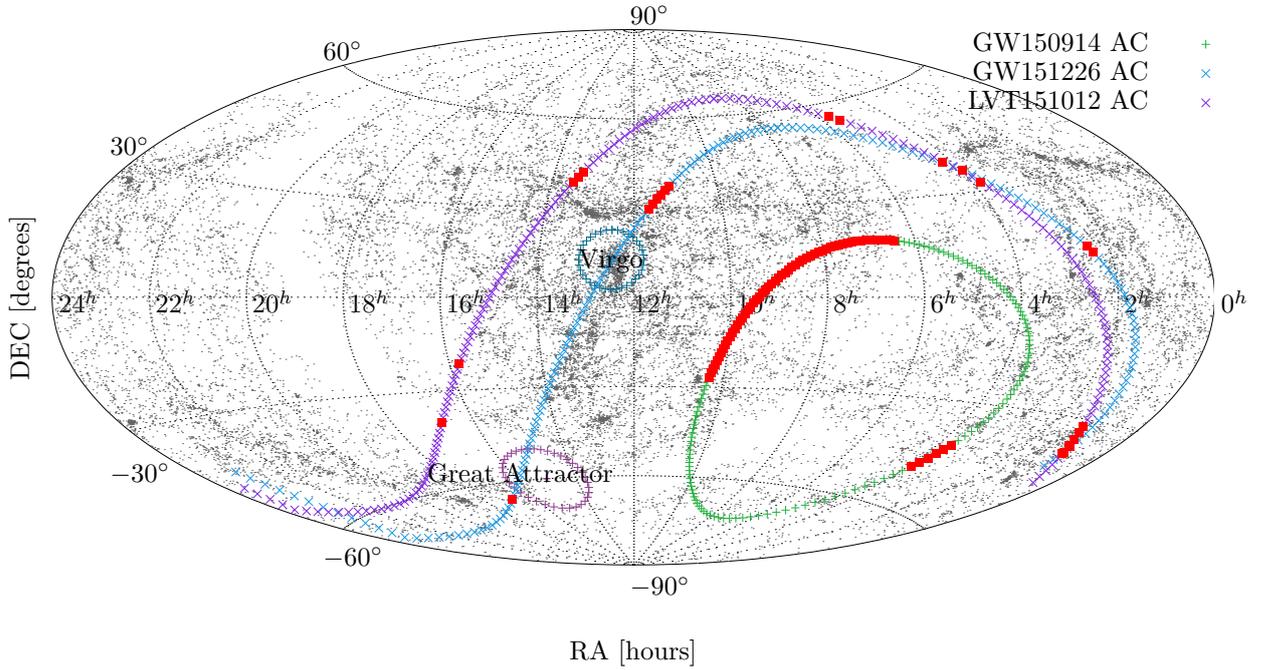

(b) for the case of a scalar (transverse or longitudinal) GW

Figure 3.6: Localization of the possible GW sources along the ACs for LIGO events GW150914, GW151226 and LVT151226, in the equatorial coordinates. The red points corresponding to the condition $G_L/G_H \approx 1 \pm 10\%$ (see (3.3.3)) represent the allowed source positions for a certain polarization state of the incoming GW.

Then for a compact binary with the almost circular orbit $e \approx 0$ in 0-PN approximation, the strain is:

$$h(t; \zeta, \Phi, \Psi) = -\frac{G\mu}{c^2 r} x^{2/3} \left( (1 + \cos^2 i) \cos 2\psi(t) F_+(\zeta, \Phi, \Psi) + 2\cos i \sin 2\psi(t) F_\times(\zeta, \Phi, \Psi) \right) \quad (3.10)$$

Obviously, in this case of mixed tensor polarization, the G-factor is more difficult to obtain than in the case of a "pure" polarization due to the incoming functional dependencies on the orbital phase $\psi(t)$, which contribute to the normalized amplitude $s(t)$ in (3.4).

Consequently, for a considered case of an orbit inclination $i$, there should be calculated the extremum of the function (3.10), which gives the maximal strain (or the wave amplitude) at the certain orbital phase $\psi_{\max}$:

$$\psi_{\max} = \frac{1}{2} \left( \frac{\beta}{\alpha} \frac{F_\times(\zeta, \Phi, \Psi)}{F_+(\zeta, \Phi, \Psi)} \right) \quad (3.11)$$

where $\alpha \equiv (1 + \cos^2 i)$, $\beta \equiv 2\cos i$ are known (or assumed).

It should be noted that the maximal orbital phase $\psi_{\max}$ is calculated for each point along the AC depending on its position in the horizontal coordinate system of the antenna.

Applying the main relation (3.5) of the method, the final formulae is:

$$\frac{h_1}{h_2} = \frac{G_1}{G_2} = \frac{\alpha \cos(2\psi_1^{\max}) F_+(\zeta_1, \Phi_1, \Psi_1) + \beta \sin(2\psi_1^{\max}) F_\times(\zeta_1, \Phi_1, \Psi_1)}{\alpha \cos(2\psi_2^{\max}) F_+(\zeta_2, \Phi_2, \Psi_2) + \beta \sin(2\psi_2^{\max}) F_\times(\zeta_2, \Phi_2, \Psi_2)} \quad (3.12)$$

which does not depend on the distance to the objects, and is determined only by the orbit inclination $i$ as well as orientation of the antennas to the source.

To sum up, the application of the method to an arbitrary tensor wave from a coalescing binary is in the following steps:

1. The construction the apparent circle (AC) for the detected event.

2. The calculation the maximal orbital phase $\psi_{\max}$ using (3.11) for the considered point along the AC for each antenna.

3. The calculation the G-factors relation using (3.12) for the considered point for the antenna couple.

4. The selecting such points along the AC where the G-factors relation corresponds to the observed strain ratio (within a margin of error) at the antennas.

The method has been applied for the LIGO events: GW150914, GW151226 and LVT151012. There is uncertainty in the calculation of the G-factors due to the unknown inclination of the orbit $i$ as well as the polarization angle $\Psi$. As a model, there were used the next two values at the $\Psi = 0°$: $i = 0°, 90°$. The results are given in Figs. (3.7) in the equatorial coordinates.

### 3.3.5 Analysis of the results. Correlations with the supergalactic plane

Besides the most common representation of the GW source localization in equatorial coordinate system, there have been considered the positions of the ACs in the galactic and supergalactic coordinates in order to detect any regularities.

The supergalactic (SG) coordinate system was introduced (see eg., Courtois et al. (2013)) for the convenience of considering the Local Super-Cluster of galaxies (hereafter LSC). The LSC has the filamentary disc-like structure with the radius $\sim 100$ Mpc, the thickness $\sim 30$ Mpc, containing galaxies clusters at the distances within $\sim 100$ Mpc (or $z \leq 0.025$). The centre of the LSC is taken to be roughly in the Virgo cluster ($SGL = 104°; SGB = 22°$), and the North Pole $SGB = 90°$ with galactic coordinates $l = 47.37°$, $b = 6.32°$ (de Vaucouleurs (1953), de Vaucouleurs (1958), di Nella and Paturel (1995), Courtois et al. (2013)).

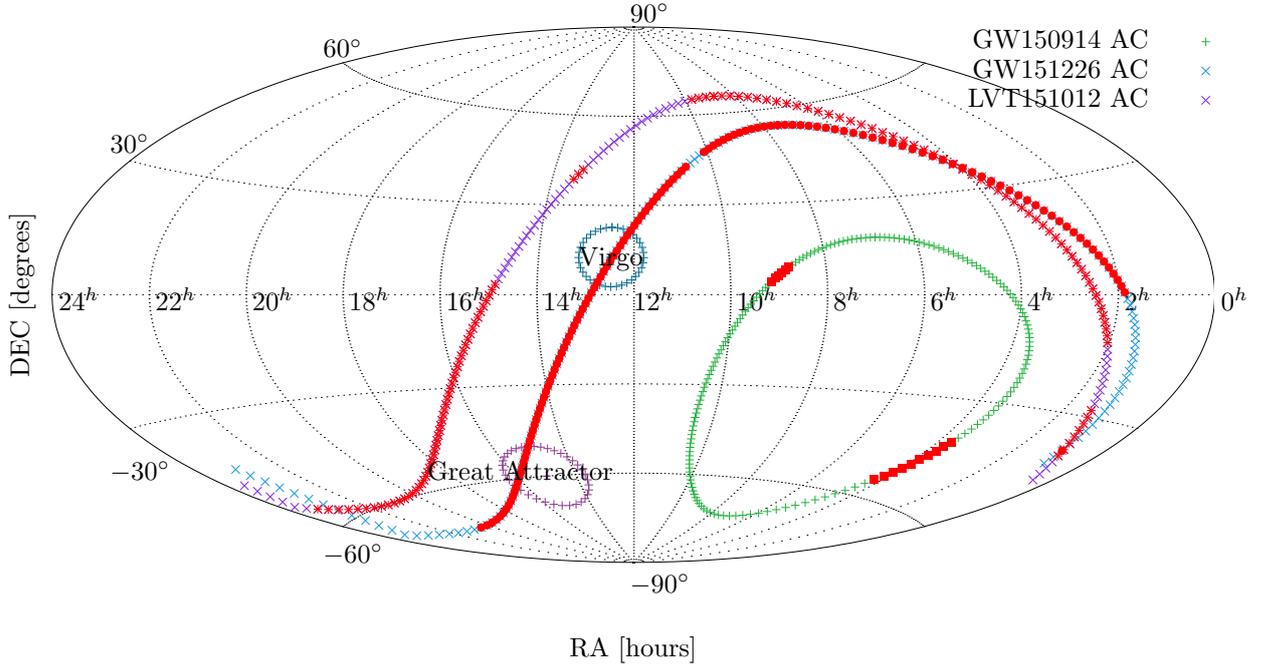

(a) $i = 0°$, $\Psi = 0°$

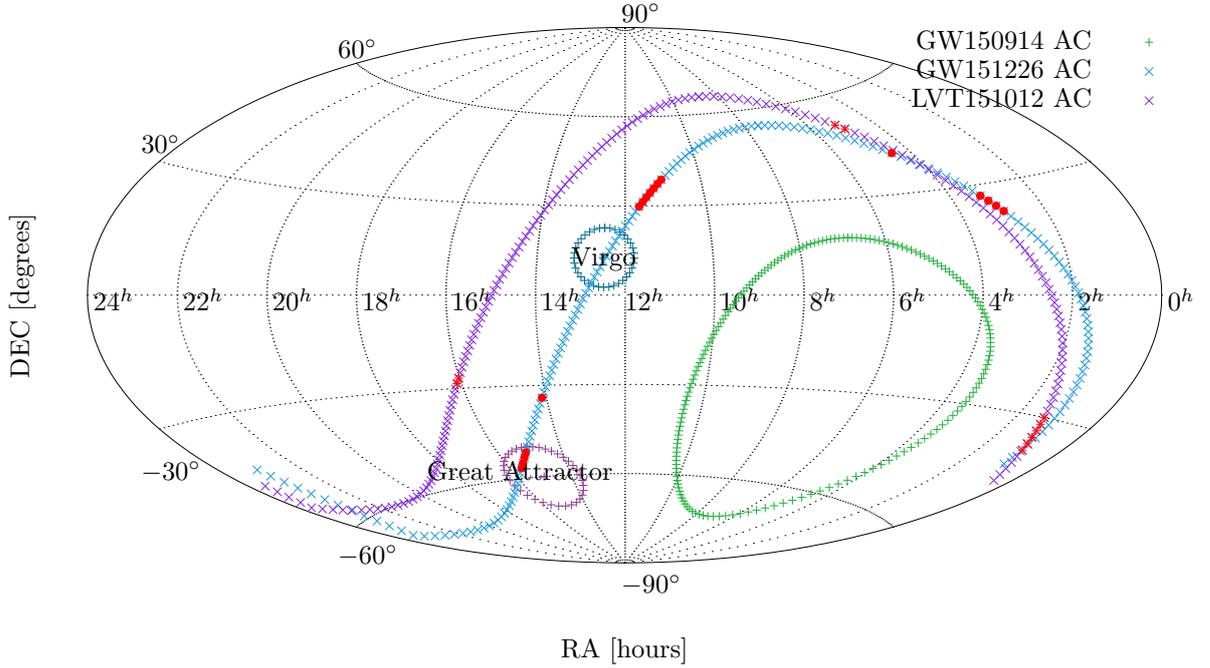

(b) $i = 90°$, $\Psi = 0°$

Figure 3.7: The source localization for the detected GW150914, GW151226 and LVT151226 in the case of tensor GW from coalescing binary with the orbit inclination $i$ and the polarization angle $\Psi = 0°$ (ch. (3.3.4)). The red points correspond to the condition $G_L/G_H \approx 1 \pm 10\%$.

In Figs. (3.8b, 3.8a) the ACs of the allowed GW sources are shown with the background projection of the 2MRS catalogue of galaxies, which is the result of the 2MASS all-sky IR survey (Huchra et al. (2012)) and includes the redshifts of 43 533 galaxies. There have been taken the subsample containing 32 656 galaxies with $z \leq 0.025$, which corresponds to the spatial distribution of the LSC.

The results of the ACs construction in SG coordinates together with the depicted GW source localization according to the method are shown on the Fig. (3.8b) for the scalar (longitudinal or transverse) and on the Fig. (3.8a) – for the tensor +" ($\Psi = 0°$) incoming GW.

Interestingly, the apparent circles for all GW events detected in 2015 lie along the supergalactic plane of the Local Super-Cluster of galaxies (Fig.(3.8a)), with a range of possible positions within $\pm 30°$ SGB.

To understand this result, a statistical model calculation was performed. There have been taken an uniform distribution (a random sampling) of the prospective sources within the region $\pm 30°$ SGB, since only by such objects can occur similar ACs. For these points, the ACs have been constructed at each hour of the sidereal time during $24^h$. The statistics in Fig. (3.9) shows the percentage of the points giving the ACs along the SG plane $\pm 30°$ SGB, per hour. It can be seen that such events can occur only in certain time intervals: $2^h - 6^h$, $9^h - 12^h$ and $20^h - 23^h$ ST, and the average percentage of these events is rougly 17.5%.

Taking an average value 0.175 as a probability that an AC will lie along the SG plane ($\pm 30°$ SGB) as well as the fact that all three LIGO events are independent, than the probability of this trend is $\sim 0.005 \approx 0.5\%$. Thereby, the statistics have shown that the position of an AC along the SG plane is quite a rare event, i.e. it may occur by a small percent of sources and only in the certain sidereal time intervals.

It is worth noting that all three LIGO events in 2015 were detected within the first time "window" (Tab. 3.2), which may witness that the possible GW sources might belong to the LSC. In this way, they would be located within 100 Mpc, which contradicts the currently accepted assumption about these sources as the binary coalescence at the distance $400 \div 1000$ Mpc (Abbott et al. (2016a)). Nevertheless, if such a correlation between the positions of ACs and the SG plane is confirmed by the forthcoming GW observations, it will also be necessary to consider alternative mechanisms for the original GW radiation. For instance, as has been discussed in the Ch. (2.3), the Feynman's field gravitation theory (FTG) predicts the existence of the scalar GW by a symmetric core-collapse Supernova (CCSN). There has been shown (Tab. 2.3) that such a source of a GW signal similar to the detected by LIGO in 2015, giving the GW with the energy $\sim 10^{-3}$ erg, may be at the distances withing the LSC (including the Virgo and the Great Attractor clusters).

This motivates one to the additional discussion about the nature of the sources of the received by LIGO signals in 2015.

## 3.4 Method for determination a GW polarization state by means of a network of interferometric antennas

In the case of a network of three and more interferometric antennas, it is possible to localize a source of an incoming GW in a unit area of the sky, where the ACs for each antennas couple intersect each other, i.e. in the case of three detectors, it is the intersection of the three ACs.

In this case, the situation changes: since the position of the source is known, there can be made the conclusions about the polarization state of the wave by means of a strains ratio on each pair of the detectors.

Consider the application of the method on the example of the event GW151226. Assuming, there were three interferometers in operation: LIGO Livingston (L), LIGO Hanford (H) and Virgo (V) (Italy), see their coordinates in the Tab. (3.1). Let the time delay between the arrivals of the signal on each antenna of the couple is known: $\Delta_{LH}, \Delta_{LV}, \Delta_{VH}$. The detection time ST of this event is given in the Tab. (3.2).

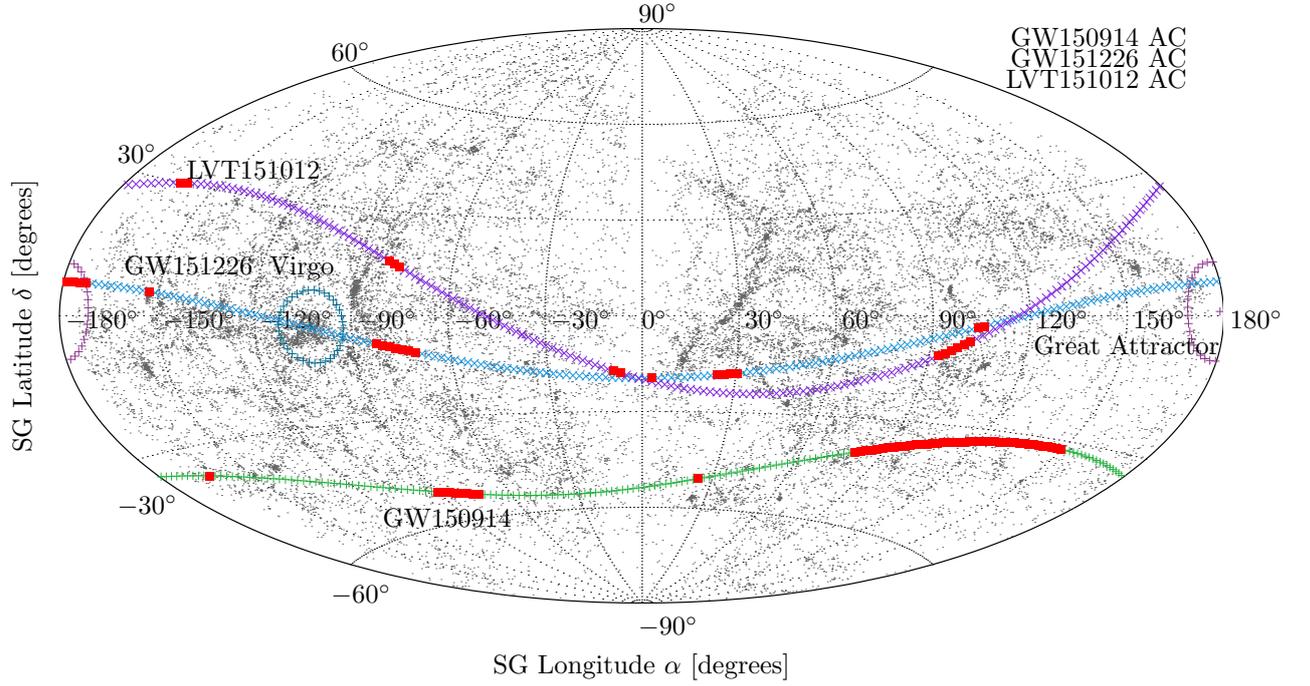

(a) for tensor "+" GW

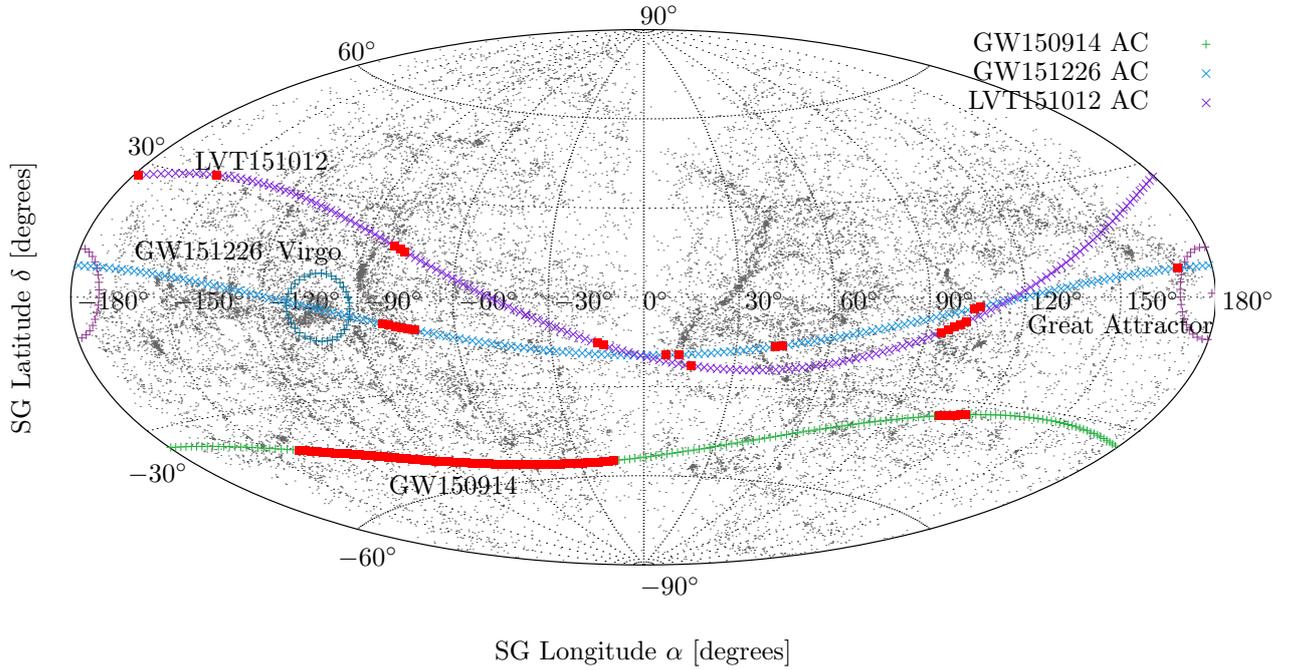

(b) for scalar (transverse or longitudinal) GW

Figure 3.8: Localization of the possible GW sources along the ACs for LIGO events GW150914, GW151226 and LVT151226, in the supergalactic coordinates. The red points corresponding to the condition $G_L/G_H \approx 1 \pm 10\%$ (see 3.3.3) represent the allowed source positions for a certain polarization state of the incoming GW.

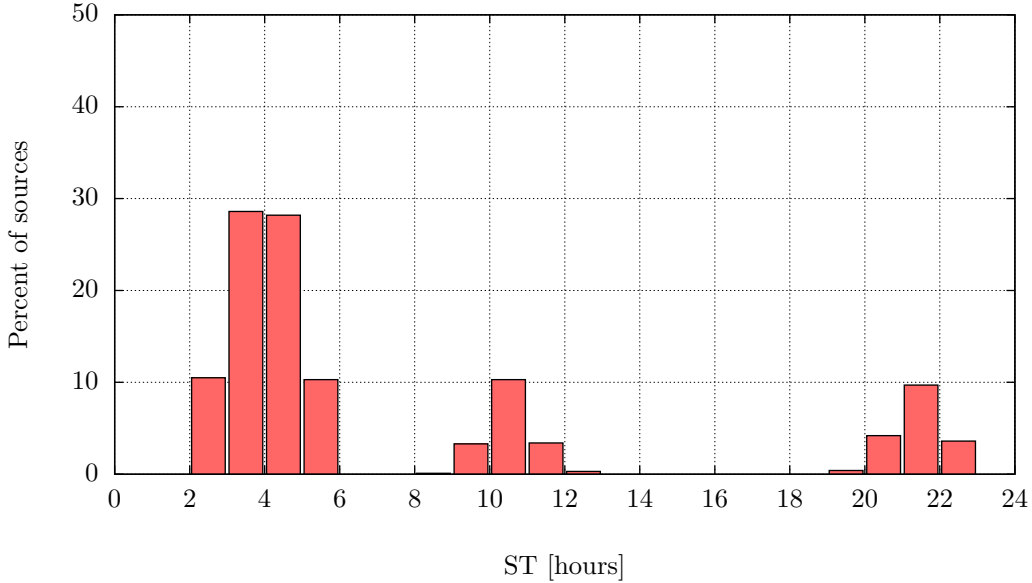

Figure 3.9: Statistics of occurrence ACs by GW sources along the SG plane within ±30° SGB.

For each pair of detectors, ACs are constructed that intersect at a certain point (with some error in the measurement of the delay), thereby determining a position of the source.

To demonstrate a possibility for distinction between different GW polarizations by means of three LIGO-Virgo antennas, there are considered two artificial sources: one located near the SG plane and another one – outside the SG plane. As a concrete test, these artificial sources were taken with the equatorial coordinates RA = $11.6^h$, DEC = $30.3°$, near the Virgo cluster, and with RA = $18^h$, DEC = $32°$ (SG $B = 70°$), i.e. outside the SG plane. The time of the artificial event has been proposed to be the same as the time of the GW151226.

The ACs for these two artificial sources are shown in supergalactic coordinates on the Figs. (3.10a, 3.10b). Red curves indicate the ACs constructed with respect to the couple LIGO Livingston – Hanford (L1-H1), blue – L1-Virgo, and green – Virgo-H1.

The G-factors for these artificial sources were calculated for each detector in the network. According to the condition (3.5), the ratio of the G-factors $G_1/G_2$ at each couple of the antennas represents the strain ratio $h_1/h_2$ to be detected at the time of GW151226, Tab. (3.3). Additionally, predictions for strain ratio at one-arm antenna clearly illustrate the possibility of recognition scalar polarization mode by means of such antenna construction.

Additionally, there are presented (Tab. 3.3) the results of the calculations for the proposed (see p. 3.1) modification of an interferometric antenna with one working arm. As has been discussed in p. (3.1), such a detector would make it possible to distinguish between longitudinal and transverse modes of a scalar wave.

To sum up, in the case when the localization of a GW source is known by means of three and more detectors in operation, it is possible to make assumptions about a polarization state of an incoming GW using the information about the detected strains ratio and matching it with the calculated ratio of the corresponding G-factors.

## 3.5 Conclusion

The proposed method for recognising possible polarization states of GWs detected by two interferometric antennas in a network provides a test on the modern gravitation theories, as well as allows us to obtain observational limitations on the parameters of physical processes resulting in the generation of powerful

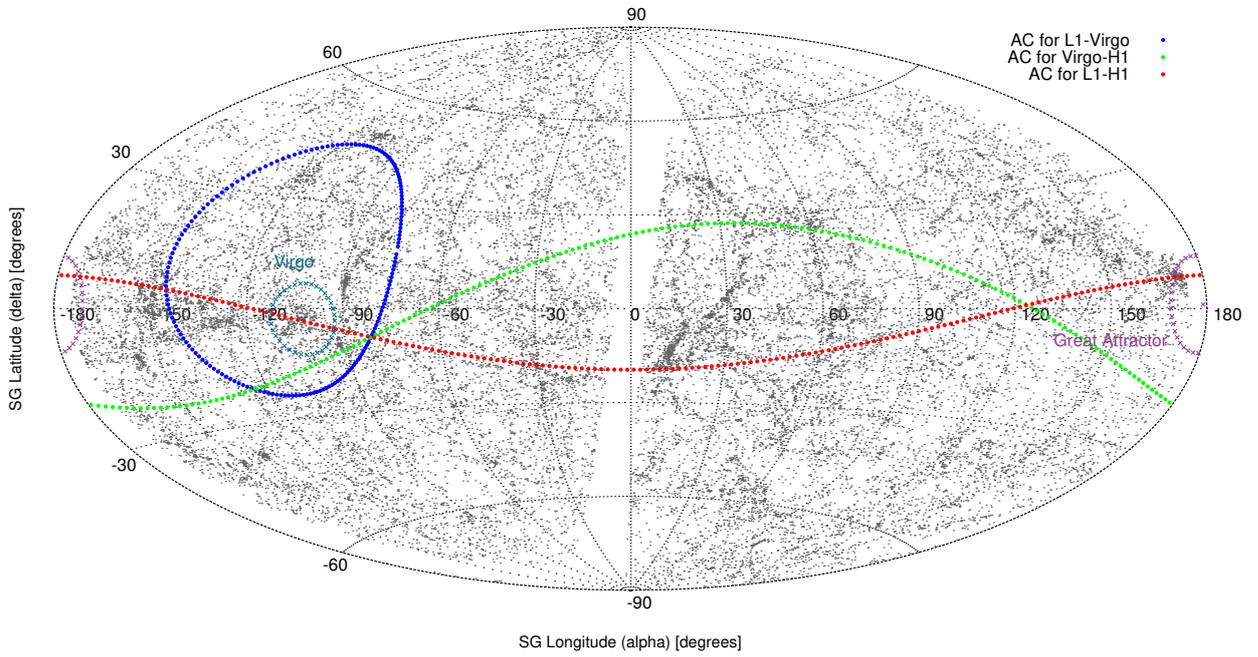

(a) The source coordinates RA = $11.6^h$; DEC = $30.3°$.

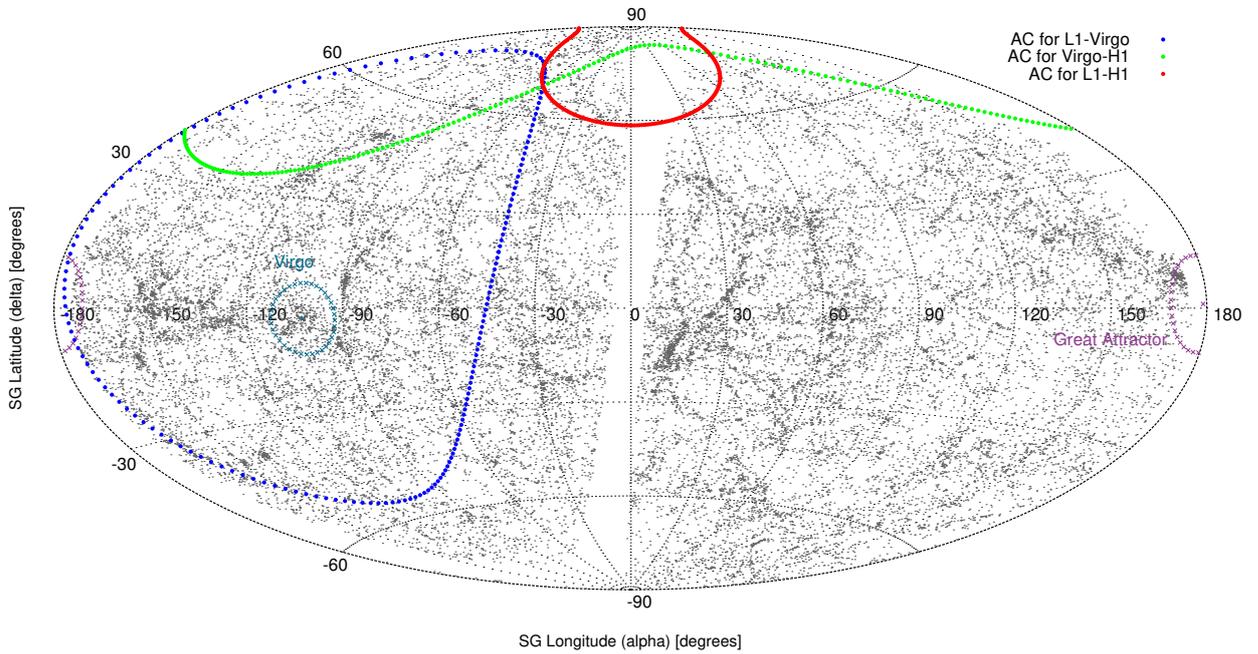

(b) The source coordinates RA = $18^h$, DEC = $32°$.

Figure 3.10: The apparent circles intersecting in the point of an artificial GW source in the case of detection by the three antennas: LIGO Livingston, LIGO Hanford and Virgo (L-H-V), at the time of the event GW151226.

Table 3.3: Calculated G-factors for two artificial GW-sources, which could be detected by LIGO Livingston, LIGO Hanford and Virgo (L-V-H) at the time of GW151226 event. The labels such as $G_{1/2}$ denote the ratio $G_1/G_2$ with respect to a couple of antennas. The time delays $\Delta$ are given in [ms].

| Polarization state | $G_L$ | $G_H$ | $G_V$ | $G_{L/H}$ | $G_{L/V}$ | $G_{V/H}$ | $\Delta_{LH}$ | $\Delta_{LV}$ | $\Delta_{VH}$ |
|---|---|---|---|---|---|---|---|---|---|
| RA = $11.6^h$ DEC = $30.3°$ | | | | | | | 1.09 | -23.35 | -0.97 |
| Tensor +, $\Psi = 0$ | 0.376 | -0.356 | -0.608 | -1.06 | -0.62 | 1.71 | | | |
| Scal. long. or trans. | 0.373 | -0.346 | -0.029 | -1.08 | -12.86 | 0.08 | | | |
| Scal. long. (1-arm) | -0.933 | 0.384 | -0.128 | -2.43 | 7.29 | -0.33 | | | |
| Scal. trans. (1-arm) | 0.359 | 0.924 | 0.992 | 0.39 | 0.36 | 1.07 | | | |
| RA = $18^h$ DEC = $32°$ | | | | | | | -9.49 | -11.53 | -25.23 |
| Tensor +, $\Psi = 0$ | -0.525 | 0.468 | 0.190 | -1.12 | -2.76 | 0.41 | | | |
| Scal. long. or trans. | -0.409 | 0.460 | 0.181 | -0.89 | -2.26 | 0.39 | | | |
| Scal. long. (1-arm) | 0.169 | 0.977 | 0.819 | 0.17 | 0.21 | 0.84 | | | |
| Scal. trans. (1-arm) | 0.985 | 0.212 | 0.574 | 4.65 | 1.72 | 2.71 | | | |

gravitational wave radiation. In the case of three and more operating detectors, the reliability of such testing significantly increases.

# Chapter 4

# Results of the follow-ups search

As has been noted, the identification of a GW signal with an EM counterpart event is critically important for both conclusions about the nature of the source and calculation of the exact distance to it, and for confirmation that the received signal is indeed a GW event from an astrophysical object. In this chapter will be analyzed the follow-ups (also called "counterparts", "transients") in the electromagnetic branch of the spectrum to the GW events GW150914 and GW151226 detected by the LIGO group. The analysis will use the results of the method (3) for the source localization in the case of the detection by the network with two interferometric antennas.

## 4.1 Motivation for the new search

Proceeding from the principles of general relativity, in which only two tensor polarization modes are predicted: "cross" and "plus", there was made the conclusion by the LIGO group (Abbott et al. (2016a)) that all sources of the events detected in 2015 are coalescing compact binaries (CBCs) at the distances $400 \div 1000$ Mpc. Therefore, the counterparts search following the registration of the GW signal was basically focused on the potential transients from such systems. At that, the search groups were oriented on the localization maps provided by LIGO (see Fig. Abbott et al. (2016b)), which initially limited the explored area and sampling of the found candidates.

At the same time, a number of modern relativistic theories of gravitation predict the presence of scalar GWs, which may be produced from spherically-symmetric core-collapse supernovae (CCSN, see discussion Ch. 2.3) over the entire range of distances where the SN bursts can occur. In other words, if the scalar radiation from such a CCSN exists, it may give a sinusoidal signal similar to that detected by LIGO, at this the source may be located at the smaller distances than the current ones estimated for the case of a CBC.

Additionally, there should be noted the proposed by Imshennik and Nadezhin (Imshennik (2010)) scenario of the explosion of a massive SN through the formation of a compact binary, which may radiate the tensor GWs, followed by the collapse into a pulsating RCO suggested to give scalar radiation.

In consideration of the foregoing, in this paper are reviewed the found by different groups candidates for EM transients to the events GW150914 and GW151226, taking into account the source localization areas according to the proposed method (3). In the introduction was briefly reviewed the possible types of the EM transients depending on the nature of the source. It is necessary to note that in the case of the gravitational core-collapse SN, there are expected the bursts of the type SNII and SNIb/c.

In this chapter, the task is to analyse the data provided by different groups for the search of the EM transients to the GW150914 and GW151226, selected according to the following criteria:

1. confidently classified supernovae SNII and SNIb/c types;

2. matching with the localization maps proposed by the method 3 for the tensor "+" and the scalar modes;

3. objects at the distances up to 100 Mpc (the LSC), $z \leq 0.025$;

4. taking into account the time interval in days between the detected GW event and the registered SN burst.

## 4.2 Searches for the transients for GW150914 and GW151226

Directly with the LIGO, a number of observatories specializing in the search for "follow-ups" in a wide range of EM spectra, as well as neutrinos, cooperate. Immediately after the receiving the GW signals GW150914 and GW151226, within 2 days after their announcement, twenty-five participating teams on the terrestrial and satellite telescopes covering the EM branch of spectra up to the 19th apparent stellar magnitude joined the search. The disseminated by LIGO probability sky maps of the source localization for the GW150914 are given in Fig. 2 Abbott et al. (2016b). The coverage area of the observations in a wide range of EM spectrum for the GW150914 is shown in Fig. 3 Abbott et al. (2016b).

Since the current generally accepted conclusion about these GW events as CBC had not yet been accepted at that moment, some of the teams searched also among the nearby galaxies, thus considering the possibility to find transients being NS-NS CBCs or CCSN. The most interesting results of these searches are discussed in this part.

### 4.2.1 Searches in the optical branch. Pan-STARRS

Search in the optical branch of spectrum was carried out using the wide-field telescope Pan-STARRS (Smartt et al. (2016)) together with the spectrographic program PESSTO specialized on finding follow-ups events. This survey covered an area of 442 sq.deg. from the provided by LIGO localization area for the event GW150914 (Abbott et al. (2016b)). As a result, 56 transients were discovered during the 41 days after the registration of the GW150914, of which 19 were classified spectrographically, i.e. they have reliable information about their nature.

From the general Pan-STARRS review (Smartt et al. (2016)), for the purposes of this work, there have been selected only the possible follow-ups identified spectrographically as SNIb/c or SNII (but not as SNIa). The corresponding sample is presented in Tab. (4.1) for the GW150914 and in Tab. (4.2) for the GW151226.

The positions of the candidates in EM "counterparts" found by the Pan-STARRS team for both GW events are shown in Fig. (4.1b) together with the possible localization of scalar-wave sources with the LSC projection from the 2MASS sample. As can be seen, these SN events in both cases are mostly within the localization taking into account the error in determining the time delay, by which the ACs were calculated. For the event GW150914, there is shown the more detail scheme in Fig. (4.2), where the AC is represented with the time delay error in ±0.5 ms. The red points mark the onfidently qualified EM counterparts being SNIb/c or SNII types, the yellow points indicate the non-identified spectrographically EM events.

Taking into account the estimates of the distances to the discussed SNs events, it should be noted that those objects that lie directly within the range of errors are much further away to give detectable "strain" GW (4.1). Thus, they should be excluded from consideration as not being the possible GW sources. The same applies to the candidates for the GW151226 (see Tab. 4.2). There are still a number of transients from the sample being within the region but not identified reliably (Smartt et al. (2016)), thereby it is difficult to make conclusions on them.

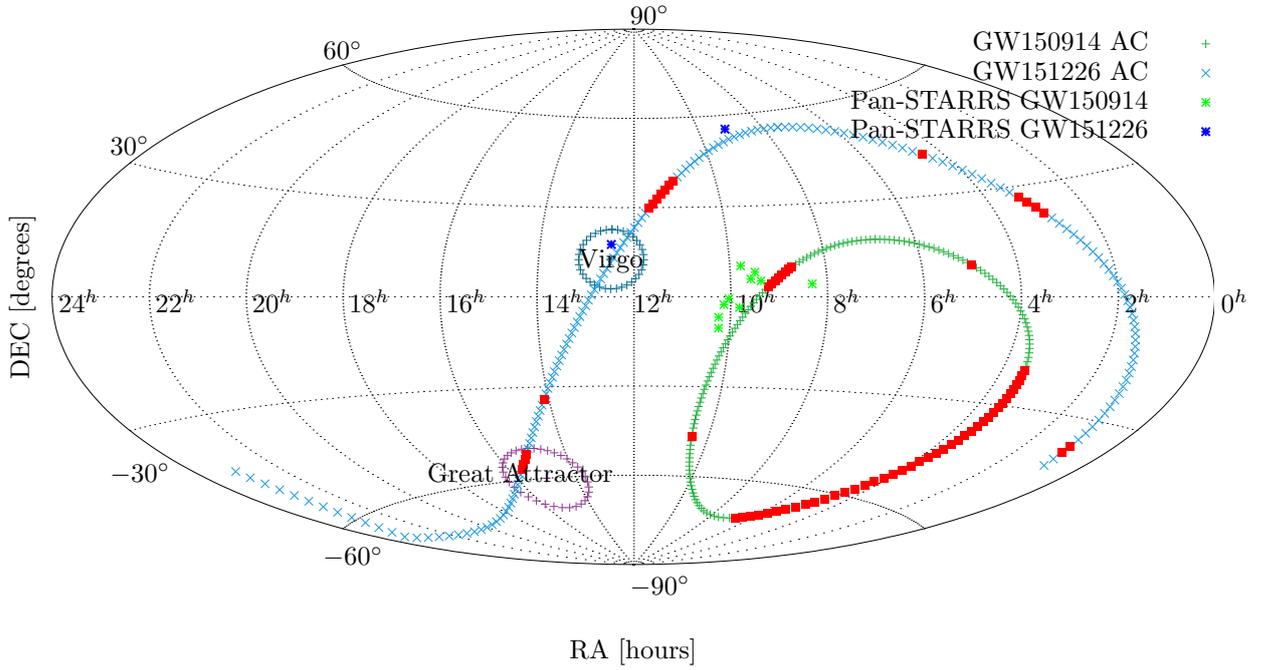

(a) for the case of tensor "+" GW

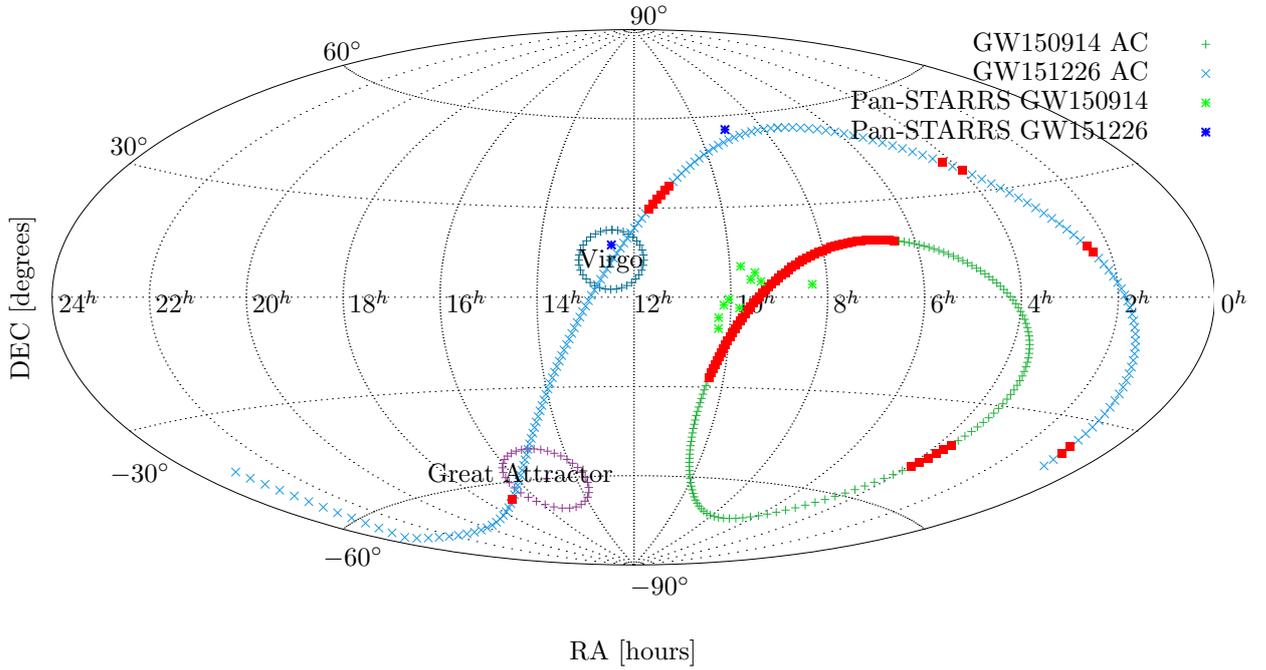

(b) for the case of scalar (longitudinal/transverse) GW

Figure 4.1: Possible localization of GW sources along the ACs for the LIGO events GW150914 and GW151226 together with Pan-STARRS optical events, in the equatorial CS. The red points represent the allowed source positions for a certain polarization state of the incoming GW calculated by the method (3) under the observed condition $G_L/G_H \approx 1 \pm 10\%$.

Table 4.1: The SNs events registered by Pan-STARRS as candidates for the GW150914 counterparts. There are given a number of days after the GW event registration in the last column.

| ID | SN type | z (redshift) | d [Mpc] | [days] |
|---|---|---|---|---|
| PS15ccx | SN Ia/Ic | 0.097/0.089 | 415/381 | 3 |
| PS15cel | SNII | 0.057 | 244 | 9 |
| PS15cki | SNII | 0.024 | 103 | 9 |
| PS15ckf | SNII | 0.019 | 81 | 19 |
| **PS15cwi** | SNII | 0.058 | 248 | 19 |
| PS15ckj | SNII | 0.020 | 86 | 19 |
| PS15cko | SNII | 0.217 | 930 | 19 |
| PS15cvy | SNII | 0.046 | 197 | 19 |
| PS15ckn | SN Ic | 0.08 | 343 | 20 |
| **PS15cmq** | SN II | 0.065 | 279 | 21 |

Table 4.2: The SNs events registered by Pan-STARRS as candidates for the GW151226 counterparts. There are given a number of days after the GW event registration in the last column.

| ID | SN type | z (redshift) | d [Mpc] | [days] |
|---|---|---|---|---|
| iPTF-15fhl | SN Ib | 0.043 | 184 | (+18 d) 2 |
| iPTF-15fhp | SN Ic | 0.129 | 553 | (+1 d) 2 |

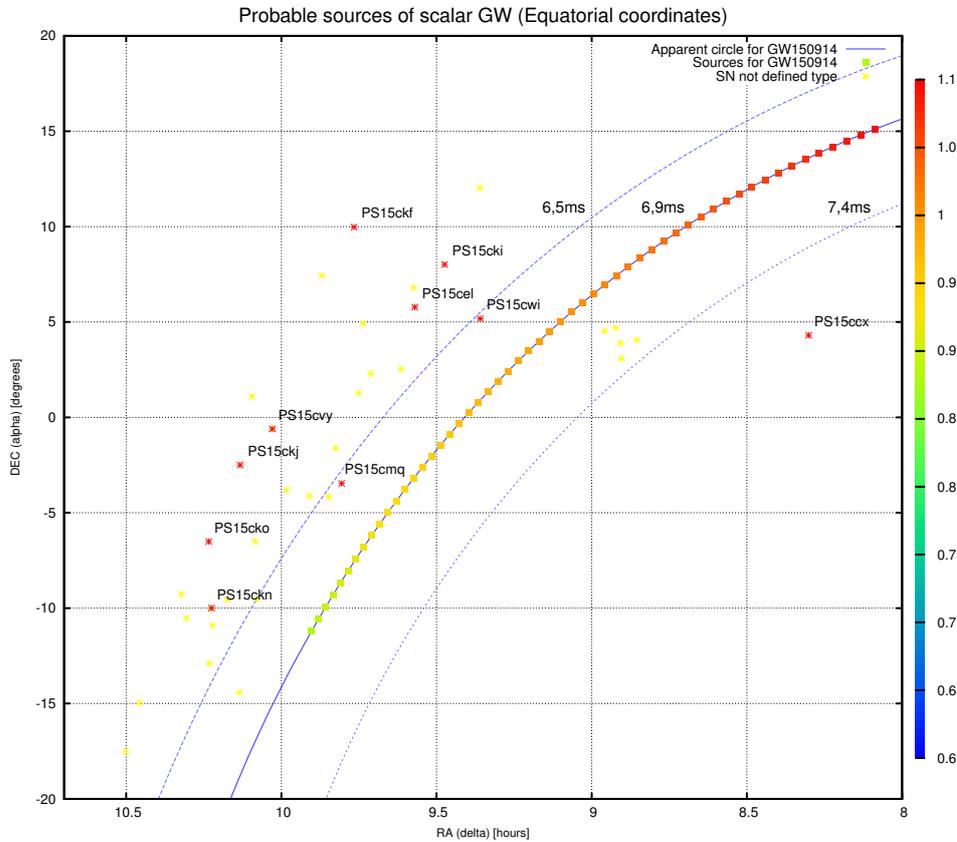

Figure 4.2: Follow-ups for the GW150914 detected by Pan-STARRS (4.2.1) in the equatorial CS. There is drawn the AC of the allowed source positions for GW150914 with the error of detection ±0.5 ms. The red points indicate the spectroscopically identified SN events, the yellow - not identified.

Table 4.3: Follow-ups discovered by the MASTER-Net in the automatic alert observations after the detection of the GW150914. DN – "dwarf-novae" (Lipunov et al. (2016)).

| N | MASTER | UTC | Type | Mag. | RA | DEC |
|---|---|---|---|---|---|---|
| 1 | J040938.68 | 16.88 | SN | 17.3 | $4^h9^m38.^s68$ | $-54°13'16.''9$ |
| 2 | J042822.91 | 16.99 | DN | 18.2 | $4^h28^m22.^s91$ | $-60°41'58.''3$ |
| 3 | J070747.72 | 21.99 | DN | 16.9 | $7^h7^m47.^s72$ | $-67°22'5.''6$ |

### 4.2.2 Observations by the global network MASTER

The global network of the robot-telescopes MASTER is a network of identical twin-tube wide-field telescopes for automatic monitoring of the optical radiation accompanying gamma- and other bursts (Lipunov et al. (2016)).

The MASTER network joined the search for optical transients for the GW150914 on September 16, 2015. Observations covered the area of 560 sq.deg. from the provided by LIGO localization area (see Fig. 2 in Abbott et al. (2016b)). The complete coverage map is given in the Fig. 2 of Lipunov et al. (2016).

From the 8 follow-ups discovered as a result of the inspection, the three were located close to the AC for the GW150914, their coordinates are given in Tab. (4.3). According to the analysis conducted by the MASTER group (Lipunov et al. (2016)), the two of the transients in question are Dwarf Novae (DN) located within our Galaxy.

The object MASTER OT J040938.68–541316.9 is very likely to be an SNI near the maximum light. Based on the difference in two days between the GW event and the recording of this SN, it was concluded that, although the SN burst belongs to the localization area, this could not be the source of the event GW150914 (Lipunov et al. (2016)).

### 4.2.3 Observation of a gamma-ray burst by the Fermi

A number of X-ray and gamma-observatories participated in the program of search for the transient of the GW150914 but only the observatory "Fermi" reported the discovery of a short gamma-ray burst (hereafter GRB) found by means of the Fermi Gamma-ray Burst Monitor Connaughton et al. (2016). This GRB duration less than 1 s occurred 0.4 s after the registration of the GW150914. The energy of the burst was estimated to be $\sim 3 \times 10^{-7}$ erg, therefore the luminosity of such an object at the distance $\sim 440$ Mpc taken by the current estimates for the GW150914 (Abbott et al. (2016b)), should amount $\sim 2 \times 10^{49}$ erg/s, which is much less than the typical isotropic luminosity of a gamma-ray burst.

There is the possibility, in the frame of the field approach (the FGT), to associate this GRB with the GW event concerning the case of the sources being a CBC comprising two RCOs without the events horizon, i.e. without infinite forces (see definition in 1.2 or the review Baryshev (2017)). In this way, there may be ejected a sufficient amount of matter to produce the burst detectable by the Fermi.

## 4.3 Conclusion

A fundamentally important conclusion of this chapter is that for the case of two operating antennas, it is necessary to search for transient events along the entire apparent circle of the allowed positions of GW sources.

Searching for the optical follow-up events are of the great importance for the gravitational-wave astronomy. Many international groups participate in the alert observations (for eg., Vlasyuk and Sokolov (2016)). At the moment, observational data are not sufficient to identify the recorded GW events with the electromagnetic transients. Therefore, it is fundamentally important, the further accumulation of optical data accompanying GW events.

# Chapter 5

# Summary and Outlook

In this paper, there were studied the polarization states of GWs predicted by the GR as well as by the scalar-tensor theories of gravity. The most promising for detection by modern interferometric antennas GW sources were considered, and relationship between physical parameters of such objects and data of a GW signal was investigated.

As a result, a new method for analysing a GW signal has been developed, which provide a possibility to recognise a polarization state of an incoming GW. This method has been applied to the GW events detected by aLIGO: GW150914, GW151226 and LVT151012. The proposed method of a GW polarization analysis may provide a new test both for physics of gravitational interaction and for astrophysical processes of a GW generation.

The main results and conclusions are the following:

1. Has been conducted the analysis of possible polarization states of gravitational waves radiated by the most promising sources in the framework of general relativity and scalar-tensor theories of gravitation (Chs. 2.2, 2.3).

2. Have been calculated parameters of a core-collapse SN model (CCSN) depending on a waveform of a radiated GW and radiated energy (Ch. 2.3).

    (a) Has been obtained the relationship between physical parameters of a CCSN and observed values of a GW signal in the case of scalar radiation in the FGT (Ch. 2.3.2).

    (b) In the case of existence of scalar radiation from CCSN, there is predicted that a GW signal, with a waveform and average parameters such as the detected by LIGO, may be emitted by an object located at the distance of 27 Mpc (within the Local-Super Cluster) and having the average radius $\sim 14.7 R_G$ and the mass $\sim 17.6 M_\odot$.

3. Has been made the analysis of response functions of an interferometric antenna on the GW polarization states, tensor and scalar (Ch. 3.1).

    (a) An interferometric antenna with two orthogonal arms can distinguish between tensor and scalar polarizations of a GW but it does not make it possible to recognise longitudinal and transverse modes of a scalar wave (p. 3.1.1).

    (b) Has been proposed a modification of an antenna with one working arm, which may distinguish between longitudinal and transverse scalar modes of a GW (p. 3.1.2).

4. Has been proposed the method for two operating interferometric antennas, which may localize a GW source on the sky depending on a polarization state of a detected GW (Ch. 3.2).

(a) There have been found possible localization areas of the GW sources for the LIGO events in 2015 with correspondence to an assumed polarization state.

   (b) For all three LIGO events in 2015 , regardless of the nature of their GW sources and polarization states, the apparent circles of the allowed positions of the sources lie within the supergalactic (SG) plane ±30° SGB (ch. 3.3.5).

   (c) Has been shown that the probability of such a parallel arrangement of apparent circles for three independent events amounts ∼ 0.5% (ch. 3.3.5).

5. Has been proposed a further method for three interferometric antennas, by means of which can be drawn conclusions about a polarization state of a detected GW (Ch. (3.4)).

6. Have been analysed electromagnetic follow-ups to the detected GW events in 2015 by the results of the searching teams: Pan-Starrs (spectroscopic), MASTER (optics) and Fermi (gamma). Have been proposed recommendations to the further search for transients (Ch. 4).